# Spatiotemporal effects of the causal factors on COVID-19 incidences in the contiguous United States

***Arabinda Maiti[a], Qi Zhang[b], Srikanta Sannigrahi[c], Suvamoy Pramanik[d*], Suman Chakraborti[d], Francesco Pilla[c]***

[a] Geography and Environment Management, Vidyasagar University, West Bengal, India

[b] Frederick S. Pardee Center for the Study of the Longer-Range Future, Frederick S. Pardee School of Global Studies, Boston University, Boston, MA 02215, USA

[c] School of Architecture, Planning, and Environmental Policy, University College Dublin, Belfield, Dublin 4, Ireland

[d] Center for the Study of Regional Development, Jawaharlal Nehru University, New Delhi, Delhi 110067, India

[*] Corresponding author: ***Mr. Suvamoy Pramanik***

Center for the Study of Regional Development, Jawaharlal Nehru University, New Delhi, Delhi 110067, India

Email: suvamo60_ssf@jnu.ac.in

## Abstract

Since December 2019, the world has been witnessing the gigantic effect of an unprecedented global pandemic called Severe Acute Respiratory Syndrome Coronavirus (SARS-CoV-2) - COVID-19. So far, 38,619,674 confirmed cases and 1,093,522 confirmed deaths due to COVID-19 have been reported. In the United States (US), the cases and deaths are recorded as 7,833,851 and 215,199. Several timely researches have discussed the local and global effects of the confounding factors on COVID-19 casualties in the US. However, most of these studies considered little about the time varying associations between and among these factors, which are crucial for understanding the outbreak of the present pandemic. Therefore, this study adopts various relevant approaches, including local and global spatial regression models and machine learning to explore the causal effects of the confounding factors on COVID-19 counts in the contiguous US. Totally five spatial regression models, spatial lag model (SLM), ordinary least square (OLS), spatial error model (SEM), geographically weighted regression (GWR) and multiscale geographically weighted regression (MGWR), are performed at the county scale to take into account the scale effects on modelling. For COVID-19 cases, ethnicity, crime, and income factors are found to be the strongest covariates and explain the maximum model variances. For COVID-19 deaths, both (domestic and international) migration and income factors play a crucial role in explaining spatial differences of COVID-19 death counts across counties. The local coefficient of determination ($R^2$) values derived from the GWR and MGWR models are found very high over the Wisconsin-Indiana-Michigan (the Great Lake) region, as well as several parts of Texas, California, Mississippi and Arkansas. The lowest $R^2$ score is found in the Northern and North-Western states (i.e., Montana, Washington, Oregon, Wyoming), Southern states (i.e., New Mexico) and North-East coast region (i.e., North Carolina and Georgia). Such associations also exhibited temporal variations from March to July, as supported by better performance of MGWR than GWR. Though the present study tried to incorporate all relevant components into the modelling, the causal impact of the other factors, such as lockdown date, the strictness of lockdown (partial or complete), restrictions on social activity and human mobility, etc. are not explored, which can be a case for future research.

| Abbreviation | Data variable | Description |
|---|---|---|
| ARSON | ARSON | Arson |
| MHHInc | Median_Household_Income_2018 | Estimate of Median household Income, 2018 |
| MHHIncPer | Med_HH_Income_Percent_of_State_Total_2018 | County Household Median Income as a percent of the State Total Median Household Income, 2018 |
| HBACM | HBAC_MALE | Not Hispanic, Black or African American alone or in combination male population |
| HBAF | HBA_FEMALE | Hispanic, Black or African American alone female population |
| DomMig | DOMESTIC_MIG_2018 | Net domestic migration in period 7/1/2017 to 6/30/2018 |
| RDomMig | R_DOMESTIC_MIG_2018 | Net domestic migration rate in period 7/1/2017 to 6/30/2018 |
| RIntMig | R_INTERNATIONAL_MIG_2018 | Net international migration rate in period 7/1/2017 to 6/30/2018 |

# 1. Introduction

The Severe Acute Respiratory Syndrome Coronavirus (SARS-CoV-2-COVID-19), first emerged in December 2019 in Wuhan province in China, has soon become a new public health concern across the world. Tracking the pathway of COVID-19 spreads across the continents, it is evident that geography and time strongly determine the virus outbreak (Fitzpatrick et al., 2020). The spillover effects and spatial transmission of the disease have been proven critical factors of the overall health burden caused by COVID-19. Spatial regression models can be useful for quantifying the risk of disease progression in the communities and developing spatially explicit maps to visualise the distribution of explanatory factors (Desmet SMU, Klaus; Romain Wacziarg UCLA, 2010; Ehlert, 2020; Xiong et al., 2020; Zhang and Schwartz, 2020). Previous studies have focused on the spatial heterogeneity of the COVID-19 transmission, considering the aggregated number of infected population (Bashir et al., 2020; Conticini et al., 2020; Sarwar et al., 2020; Xiong et al., 2020; Yao et al., 2020). Developing spatial models and understanding the confounding effects of the variables is critical to sense the spatial variation of virus transmission at the local scale (Mollalo et al., 2020; Ren et al., 2020; Zhang and Schwartz, 2020).

Studies have utilised environmental, social, economic, demographic variables to explain the spatial variability of the COVID-19 incidents and discovered the underlying risk of the outbreaks across scales (Karaye and Horney, 2020; Qi et al., 2020; Ren et al., 2020; Sannigrahi et al., 2020b). Fortaleza et al., (2020) noted trade-offs between travelling distance and new incidence of the COVID-19 cases in Sao Paulo city, Brazil. Bolaño-Ortiz et al., (2020) found that particulate matter ($PM_{2.5}$) has strongly associated with the virus spread in Brazilian cities. Almagro and Orane-hutchinson, 2020; Borjas, (2020) documented a strong association between the ethnic composition and COVID-19 cases in New York City of United States (US).

These studies (Almagro and Orane-hutchinson, 2020; Borjas, 2020) observed that neighbourhoods of the Hispanic population with high household density are likely to be most infected by COVID-19. Thakar (2020) explored the occurrence of community transmission of COVID-19 to the nearby locations and recommended to impose a strict restriction on community mobility and social distancing to curb the virus spread in its initial stage. In addition, other studies have also focused on the measures of mobility (Oztig and Askin, 2020), migration (Chen et al., 2020; Xiong et al., 2020) and air transport (Christidis and Christodoulou, 2020), which could reduce the spillover effects of virus out breaks.

Several studies have explored the spatial interaction among the COVID-19 fatalities and socioeconomic, demographic and environmental factors. Ren et al., (2020) utilised the ecological niche model to identify the potential risk zone based on the socioeconomic data in three metropolitan cities of China. The purpose of study (Ren et al., 2020) was to develop an early warning system, by which spatial interaction among the confounding factors could be observed in details. You et al., (2020) study in Wuhan found that population density, proportion of construction land, aged population density, tertiary industrial output per unit land, can upsurge the COVID-19 morbidity. But, the cross-country comparison of virus spread and their interaction with demographic, economic and environmental parameters are limited. Among them, Sannigrahi et al., (2020b) focused on the European region, and carried out the spatial regression model to understand the spatial heterogeneous association among the factors in different European countries. This study (Sannigrahi et al., 2020b) found that demographic variables have the highest influence on COVID-19 fatalities in Germany, Austria, Slovenia etc. Ehlert, (2020) had observed similar association in Germany. Iyanda et al., (2020) utilised Multiscale Geographical Weightage regression to understand retrospective phenomena of cross-national virus outbreaks. Similarly, Karaye and Horney, (2020); Mollalo et al., (2020); Sun et al., (2020); Zhang and Schwartz, (2020) studied spatial variability and social vulnerability COVID-19 risk in cross-county analysis in the US. Mollalo et al., (2020) considered 90 days aggregated data of COVID-19, and selected four explanatory variables from the thirty-five explanatory variables covering socioeconomic, behavioural, environmental, topographic and demographic indicators in order to develop a spatial regression model. Desmet SMU, Klaus; Romain Wacziarg UCLA, 2010; Sun et al., (2020) highlighted the spatial effects of time-variant explanatory variables on the overall COVID-19 incident in the US. (Pourghasemi et al., 2020) focused on the cross country spatiotemporal spillover of the virus spreads in Iran and has not considered the spatial interaction of socioeconomic and demographic factors. Zhang and Schwartz, (2020) analysed spatial disparities across US counties, and identified the heterogeneity of the virus spread in the metropolitan and non-metropolitan areas through a global spatial regression model.

Omitting the time variable in spatial models can lead to erroneous estimates and misleading conclusion. Although several researches have employed various spatial models (both global and local) in COVID-19 studies, there has not been substantial effort to capture the time effect in spatial regression modelling. Moreover, assuming the time-independent and homogenous impact of the confounding factors on response variables (COVID-19 cases and deaths in the present case) might introduce ambiguity in parameter approximation and eventually produce unconvincing results, which should not be expected for a sensitive case of experiment such as COVID-19. Therefore, the present research has made an effort to address the current research gap in spatial COVID-19 studies by conceptualising many time-dependent spatial regressions models using open source data information in the contiguous United States (US). The **hypothesis** of this study is framed as "*the spatial interactions between the confounding factors and COVID-19 counts strongly depend on time, thus space entity alone cannot fully explain this association*". The specific objectives of this study are **1**. to explore the overall association between the explanatory factors and COVID-19 cases and deaths; **2**. to examine the local interactions between the explanatory factors and COVID-19 incidences; **3.** to develop dynamic spatial regression models for exploring the time-dependent local spatial association between the explanatory and response variables; **4.** to measure the relative importance of variables for developing parsimonious regression models.

# 2. Materials and methods:

## *2.1 Data source and pre-processing*

The present research has utilized the county-level most updated aggregated datasets provided by John Hopkins University (Killeen et al., 2020). The mentioned datasets contain 348 relevant variables covering all domains, such as demography, education, economy, health care capacity, crime statistics, public transit, climate, and housing. The details of the 348 variables are given in **Table. S1**. Since the main aim of the present study was to establish a modelling framework to examine the space, and time-dependent association between COVID-19 counts and causal factors, all the relevant variables considered in this study were pre-processed to connect the values of each variable to their corresponding county unit, which is identified by its unique Federal Information Processing Standard (FIPS) code. In the five-digit FIPS code, the first two digits refer state information, and the last three digits describe county information. The John Hopkins team has been retrieving information from various governmental and institutional sources that include United States Census Bureau, United States Department of Agriculture (USDA) Economic Research Service, the National Oceanic and Atmosphere Administration (NOAA), the Association of American Medical Colleges (AAMC), Henry J. Kaiser Family Foundation (KFF), the Center for Neighborhood Technology (CNT), the Bureau of Justice Statistics, and Department of Justice (DOJ) (Killeen et al., 2020).

The cited data has also retrieved key information on the health care system at county scale to examine how a county with better and improved health care system handling COVID-19 efficiently and vice versa.

The daily COVID-19 counts, including confirmed cases and deaths, were collected for 22 January to 26 July 2020 period from USAFacts (https://usafacts.org/visualizations/coronavirus-covid-19-spread-map/). The daily counts of COVID-19 cases and deaths were converted to cumulative sum for further analysis and subsequent interpretation. The USAFacts team aggregates the most updated COVID-19 counts from various sources, including Centers for Disease Control and Prevention (CDC) and state- and local-level public health agencies. Following, the county-level COVID-19 data was again verified by communicating state and local agencies directly. However, for most of the states, the USAFacts team gathers the daily county-level cumulative COVID-19 counts (positive cases and deaths) from the published table, web dashboards, or PDF report available on state public health website through scraping or manual entry. The actual numbers (COVID-19 counts) reported in the USAFacts report sometimes may not exactly match with the state public health organisation reports and statistics. This could be due to the frequency in which the USAFacts are collecting and updating data is different from the data collection frequency of local governmental organisations. Additionally, there are a few unique states where up to date the county-scale data is either not available on public health website, or data collection is not frequent. For example, the updated COVID-19 counts in California and Texas state are not available on the state public health websites. For these sates, the USAFacts team has extracted the latest available numbers from county-specific public health websites.

Daily air pollution data was collected from OpenAQ data repository system for five key air pollutants, particulate matter ($PM_{2.5}$, $PM_{10}$), Nitrogen Dioxide ($NO_2$), Carbon monoxide (CO), and Sulfur Dioxide ($SO_2$). The daily concentrations of these air pollutants were converted to the monthly average unit to examine its association with COVID-19 casualties. So far, OpenAQ platform consists of 686 million air quality measurements, 150 data sources, 13000 locations, and 95 countries in their system that enabling the system to collect hourly air pollution concentration estimates from governmental and sensor sources. The "*ropenaq: Accesses Air Quality Data from the Open Data Platform OpenAQ*", an R package, was utilised for accessing the large volume of data for the entire US from 22 January to 27 July 2020. The location wise air pollution data was further converted to raster surface using IDW interpolation method. Following, the average air pollution concentration of each county was calculated using zonal statistics as table function.

### *2.2 Variable selection and dimensionality reduction*

Dimensionality reduction and extracting critical information from datasets is crucial for regression modelling and effective decision analysis. The present research has employed a stepwise forward regression approach as a tool for separating the key variables from the sets of unorganised variables. A total of seven groups, i.e. crime, demography, education, employment, ethnicity, health, and migration, which were assumed to have both synergistic and tradeoff association with COVID-19 counts, were formed, and subsequently, the key variables were extracted from each group on the basis of variable inflation factor (VIF) and model variability score, characterised by the coefficient of determination ($R^2$) and adjusted coefficient of determination (Adj. $R^2$). During the initial screening, it has been detected that data was not consistent for all the 348 variables considered, and therefore, the final stepwise regression modelling was performed considering the filtered variables. For crime category, total 16 variables have been incorporated into the modelling, while for the other category, a total of 14, 29, 6, 72, 63, 5, 7, and 4 variables were considered for demographic, education, employment, ethnicity, healthcare, pollution, migration, and climate, respectively (**Table. S2**).

Multiple collinearity tests, including VIF, $R^2$ change, correlation coefficient, probability, and t statistics, were executed to detect the redundant variables in the models. High collinearity would be evident in the model if the VIF value is found >10; therefore, all the filtered variables considered in the regression modelling were scrutinised to eliminate the redundancy in model parametrisation. Followed by stepwise forward regression, the enter stepwise regression method was performed to measure the VIF value of each explanatory variable in order to ensure that the multicollinearity was entirely eliminated. The final parsimonious models that relied on fewer parameters and at the same time explained the maximum model variances with less uncertainty were parameterised for each category for both COVID-19 cases and deaths. It should be noted that the confounding effects of the independent variables have not been assessed thoroughly, which could be a scope for future research.

## *2.3 Spatial regression*

### 2.3.1 Global Spatial Regression models

Spatial regression models have been used extensively in COVID-19 research across the scale (Guliyev, 2020; Chakraborti et al., 2019; Mollalo et al., 2020; Sannigrahi et al., 2020b; Sannigrahi et al., 2020a; Sun et al., 2020; You et al., 2020). Among the available global spatial regression models, we used Ordinary Least Square (OLS), Spatial Error Model (SEM), and Spatial Lag Model (SLM) for measuring the global associations between causal factors and COVID-19 counts at the county scale. The OLS model, which usually assesses the interactions between explanatory variables and dependent variable (COVID-19 cases and deaths), was conceptualised as follows:

$$y_i = \beta_0 + \beta x_i + \varepsilon_i$$

Where $y_i$ is the COVID-19 case or death counts at county $i$, $\beta_0$ is the model intercept, $\beta$ is the slope of the regression coefficient, $x_i$ is the selected independent variable(s) at county i, $\varepsilon_i$ is the error term at model estimates. The global OLS assumed to have spatial stationarity across the scale, and therefore, also hypothesised that a model conceptualised for a particular area can be applied effectively to other areas of interest (Fang et al., 2015). According to Anselin and Arribas-Bel, (2013), the global OLS is based on two fundamental assumptions, first, the observation in the feature space does not vary with space and therefore should be independent in nature, and second, the residual model errors should not be correlated (Oshan et al., 2020; Mollalo et al., 2020).

The spatial lag model (SLM) assumed spatial dependency between the explanatory and response variables in feature space and conceptualised the global regression by incorporating spatial dependence attributes in the modelling. The SLM also assumes to have spatially lagged dependent variable in the modelling, which can be ensured by the spatial dependence test resulted from OLS. If the determinant factors, including Moran's I (error), Lagrange Multiplier (lag), and Robust LM (lag), exhibited statistically significant estimates at a defined probability level, then one should reconsider the model selection process and go for SLM as a replacement for OLS. The SLM can be defined as:

$$y_i = \beta_0 + \beta x_i + \rho W_i y_i + \varepsilon_i$$

Where $\rho$ is the spatial lag component, $W_i$ is spatial weights (spatial weights matrix in a row format). Spatial weight matrix was generated using multiple approaches, including contiguity-based methods (Queen contiguity and Rook contiguity) and distance-based methods (Euclidean distance, Arc distance, and Manhattan distance). Contiguity-based weight was approximated using first order of contiguity. The county unique identifier number was utilised as a base for weight calculation. Since the accuracy and performances of all the global regression models strongly rely on spatial weights, we adopted both contiguity and distance-based weights for comparing the results at varied parameter setups.

The spatial error model (SEM) is an extension of global models that fundamentally stands on the assumption of spatial dependence in the residual error of OLS (Fang et al., 2015; Song et al., 2014; Chi and Zhu, 2008; Yang and Jin, 2010; Guliyev, 2020; Sannigrahi et al., 2020c ; Mollalo et al., 2020). The SEM assumes that spatial autocorrelation among the regression residuals is thus evident and two standard spatial dependence tests, such as Lagrange Multiplier (error) and Robust LM (error), were executed to ensure the statistical significance in spatial dependency in error terms, specified as follows.

$$y_{it} = x^{i}_{it}\beta + u_i + \varepsilon_{it}$$
$$\varepsilon_{it} = \lambda W_{\varepsilon t} + v_{it}$$

Where $\lambda W_{\varepsilon t}$ is the spatial error term, $\lambda$ denotes the autoregressive factor, $v_{it}$ refers to the random error term, that is normally conceptualised to be independent and ideally distributed in feature space, $\varepsilon_{it}$ refers to the spatially uncorrelated error term (Guliyev, 2020). The SEM model consists of two error terms, i.e., $\lambda W_{\varepsilon t}$ and $\varepsilon_{it}$. Since the OLS derived spatial dependence test has suggested a statistically significant spatial dependency among the observations for SLM and SEM, this study considered all the three standard global spatial regression models for spatial modelling and subsequent interpretation. The results of the spatial dependence test showed that both LM (lag and error) and Robust LM (lag and error) exhibited the statistical significance estimates; therefore, both SEM and SLM were utilised to assess the synergies and tradeoffs between COVID-19 counts and associated factors at the county scale. When estimating the global models, both dependent and independent variables were converted to cumulative sum units. Additionally, the global association between the variables were assessed for all the seven sub-components for capturing the individual effect of each sub-component on COVID-19 counts at the spatial feature space.

### 2.3.2 Local Regression

In many real-life cases, the general global assumptions and spatial stationarity among the observations in feature space, could be ineffective and thus produced inelastic and biased estimates at the local scale. Since the main objective of the present research is to establish predictive spatial models at the local scale, two most used local spatial regression models, i.e. geographically weighted regression (GWR) and multiscale geographically weighted regression (MGWR), were employed for local spatial regression modelling and subsequent interpretation (Sannigrahi et al., 2020d; Oshan et al., 2019). The GWR model is developed on the basis of Toddler's first law of geography, which suggests "everything is related to everything else, but near things are more related than distant things." In GWR, each observation in feature space will vary over space and therefore, would be associated with locally varying coefficients of the regression parameters. This addition of local spatial context in GWR modelling favours exploring the spatial dependency among the parameters. GWR can be defined as:

$$y_i = \beta_{i0} + \sum_{j=1}^{m} \beta_{ij} X_{ij} + \varepsilon_i$$
$$i = 1,2.\dots.,n$$

Where $y_i$ is dependent variable (COVID-19 case or death counts) at county i, $\beta_{i0}$ refers to the regression intercept, $\beta_{ij}$ refers to the independent regression parameter, $X_{ij}$ is the value of the jth regression parameter, $\varepsilon_i$ refers to the regression error.

Although GWR models have been embraced as a solution for global spatial stationarity in regression estimates, the same has been suffered in cases when a constant and straightforward bandwidth has not been able to detect the spatial non-stationarity at varying bandwidths across the feature space. To address this problem, Fotheringham et al., 2017; Oshan et al., 2019, proposed a multiscale and multi bandwidth GWR, that allows exploring the local relationships among the factors at varying spatial scales by computing shifting bandwidth based on the distributions of observation. MGWR can be defined as:

$$y_i = \sum_{j=0}^{m} \beta_{bwj} X_{ij} + \varepsilon_i$$
$$i = 1, 2 \ldots .., n$$

Where $\beta_{bwj}$ refers to the differential bandwidth at feature space. The rest is the same as discussed in GWR.

## *2.4 Variable Importance*

Machine Learning models have been used extensively in measuring feature importance in multiparameter models. The present research utilised the supervised machine learning algorithm, Random Forest (RF), for spotting the key causal explanatory factors in the models. RF models (Breiman, 2001; Chakraborti et al., 2020), fundamentally based on bootstrap aggregating of decision trees, can minimise the unexplained variance of models and thus improve prediction accuracy (Altmann et al., 2010). RF models have been utilised for many domain-specific studies, including, gene expression-based cancer classification (Okun and Priisalu, 2007), biology of ageing (Fabris et al., 2018); remote sensing land cover mapping (Ma et al., 2017); screening underlying lead compounds (Cao et al., 2011); Structure damage detection (Zhou et al., 2014), etc.. In this study, we measured the variable importance based on the overall capacity of the variables to explain the total model variances. Relative Importance (RI) and normalise importance scores were also computed for each variable to verify the predictive accuracy of the models as well as individual contribution of each variable to the overall model performances.

### *2.5 Experimental design*

In this study, we structured the entire analysis into a few sequential and logical steps to carry out the entire analysis. The global and local spatial regression analysis has been carried out through four separate models:

*Model 1: Global regression model built on considering both static dependent and independent variables*

Model 1 was conceptualized for doing global regression analysis between COVID-19 counts and explanatory factors. The daily COVID-19 data from 22 January to 26 July was converted to cumulative sum for changing the nature of the data from dynamic to static. Only the final filtered variables for COVID-19 cases and deaths were considered for model 1. Group-wise assessment was not done for model 1. The final selected variables, 6 for cases and 6 for deaths, had exhibited acceptable VIF scores, and therefore, multicollinearity problems in the model did not seem evident for all the multi-parameters regression models. All the global models, including OLS, SEM, and SLM, were conducted using GeoDa and GeoDa Space software. The first order Queen and Rook contiguity was applied for spatial weight estimation. The distance-based approach was utilised for generating the spatial weights of the observations. In particular, the Euclidian distance method was adopted for distance-based spatial weight calculation.

*Model 2: Local Regression model built on using static independent and static dependent variables*

Model 2 was developed by incorporating both static independent and static dependent variables into the modelling. Local GWR and MGWR modelling was undertaken to explore the local correlation and association between the explanatory and response variables. Both GWR and MGWR were performed in the MGWR software package, developed by Olson et al., 2019. For model 2, only the final filtered variables (6 for cases and 6 for death) were taken as independent variables. Using these variables, seven parameters local regression models have been developed for COVID-19 cases and death. The cumulative sum values of COVID-19 cases and deaths were accounted for model 2.

*Model 3: Group-wise local regression model built on using static independent and static dependent variables*

Model 3 was conceptualised after incorporating group-wise (crime, demography, education, employment, ethnicity, health, and migration) variables into the modelling. Using the stepwise forward and enter regression method, the filtered variables (VIF<4) for each group was identified. Among the seven major groups, a total of two variables (county population agency report crimes and ARSON), one variable (female age 85+), two variables (less than a high school diploma 2014-18 and bachelor's degree or higher 2014-18), three variables (unemployed 2018, median household income 2018, Median household income percent of state total 2018), two variables (HBAC_MALE and NH_FEMALE), two variables (Geriatric Medicine and Preventive Medicine), and a total of three variables (Population estimate 2018, domestic migration 2018, and R international migration 2018), have been selected for crime, demography, education, employment, ethnicity, health, and migration for developing local regression models for COVID-19 cases. Similarly, for COVID-19 deaths, a total of two variables for crime (Robbery, Motor vehicle thefts), one variable for demography (female age85+), one variable for education (bachelor's degree or higher 2014-18), three variables for employment (unemployed 2018, median household income 2018, median household income percent of state total 2018), two variables for ethnicity (HBA Female, BA Female), one variable for health (endocrinology diabetes and metabolism specialists (2019)), and four variables for migration (Pop estimate 2018, domestic migration 2018, R international migration 2018, and R domestic migration 2018), have been considered for COVID-19 deaths.

*Model 4: Dynamic local regression model built on using dynamic dependent and static independent variables*

In model 4, the monthly COVID-19 cases and deaths were chosen as dependent variables, and the annual averaged static group variables were considered as independent variables. The monthly sum values of COVID-19 cases and deaths were calculated for March, April, May, Jun, and July. A total of ten (five for cases and five for death) multiparameter local spatial regression models were developed for exploring the dynamic association between the response and explanatory factors. The final filtered variables (six for cases,  ARSON, median household income 2018, median household income percent of the state total 2018, HBA male, domestic migration 2018, R international migration 2018) and six for deaths, i.e., median household income 2018, median household income percent of state total 2018, HBA Female, domestic migration 2018, R international migration 2018, and R domestic migration 2018) was incorporated for the dynamic local regression modelling.

The adaptive bisquare spatial kernel weighted method was employed for approximating the kernel bandwidth for GWR and MGWR models. The default golden bandwidth search approach was chosen for computing uniform (GWR) and locally varying (MGWR) bandwidths. Among the different optimisation criteria, i.e., AICc, AIC, BIC, and CV, the AICc approach was considered for selecting optimal bandwidth in the feature space. Local correlation diagnostics, including, condition number (CN), local spatial VIF, local variance decomposition proportions (VDP), etc. have also been computed for evaluating the local collinearity among the observations and parameters. Bandwidth confidence intervals were also measured at different probability levels to ensure the reliability of spatially varying bandwidths, derived from the MGWR.

# 3 Result

## *3.1 Spatial pattern of COVID-19 cases and deaths in the contiguous US*

Spatial distribution and pattern of COVID-19 cases and deaths per 10,000 people in the contiguous United States (US) is shown in **Fig. 1**. Multiple spatial clusters of simultaneously high numbers of cases and deaths were formed, which designate the unequal and heterogeneous distribution of COVID-19 counts across the counties. Among the clusters, four main clusters have been identified throughout the study period. The first cluster was formed over the North-Eastern coastal region, covering Massachusetts, Connecticut, Pennsylvania, Washington D.C., Marry Land, New Jersey and part of New York (New York City in particular). The second cluster was observed in the South-Eastern region, which covers Mississippi, Alabama, Georgia, South Carolina, North Carolina, and Florida. The third cluster was found in the great lake region – Michigan, Wisconsin, Illinois, centered at Chicago (one of the largest cities in the country) of Illinois. Last, the South-Western region including southern California, Arizona, New Mexico (northwestern part), and Colorado was also among the areas with high numbers of cases and deaths (**Fig. 1**).

## *3.2 Association between explanatory factors and COVID-19 cases/deaths*

### 3.2.1 Model 1: Static global regression analysis

Three global regression models, OLS, SLM and SEM, were performed for examining the global and spatial non-stationary association between the explanatory factors and COVID-19 case and death numbers (**Table. 1**).

For COVID-19 cases, the coefficient of determination ($R^2$) statistics, which denote the overall model strength and robustness, were measured as 0.78, 0.80 and 0.80 for OLS, SLM and SEM, respectively. The spatial dependence diagnostics criteria for the OLS model, i.e., LM Lag and LM error, were found statistically significant, thus indicating the requirement of more appropriate and relevant global models, such as SLM and SEM **(Table S2)**. The AIC value, which denotes the overall model accuracy and parsimonious character of the models, was shown to be the lowest (most relevant) for SLM, followed by OLS and SEM. This suggests that the SLM model could be the most relevant global regression model that has explained the maximum model variability of the model. Regarding the correlations of the explanatory variables[1], ARSON, MHHInc, MHHIncPer, and HBACM were positively correlated with the number of COVID-19 cases. Among these four covariates, HBACM was found to have the most statistically significant relationships with the number of cases, given that its t-statistic/z-statistic was the highest in each model (52.79, 39.3 and 35.04 in OLS, SLM and SEM, respectively). The next with substantial significant coefficients was ARSON with its t-/z-statistics being 14.88, 21.2 and 23.27 in OLS, SLM and SEM, respectively. MHHInc and MHHIncPer were found to have smaller significance values, with the former is statistically significant (at the 5% level) only in SLM and the latter is statistically significant (5% level) in OLS and SEM. Meanwhile, DomMig was found negatively (significantly) correlated with the COVID-19 cases in all the three models. Last, RIntMig shows statistically insignificant associations with the cases, although its results were different across the models.

Moving onto the number of COVID-19 deaths, the $R^2$ values were 0.36, 0.69 and 0.69 for OLS, SLM and SEM, respectively. The AIC value was found lowest in SLM, compared to those in OLS and SEM, indicating that the SLM model would perform better under the given modelling framework. To interpret the explanatory variables, DomMig, RIntMig and RDomMig were significantly associated with the number of deaths for all three models and their associating directions were consistent. Specifically, RIntMig and RDomMig covariates were found to be positively correlated with deaths, while DomMig (the one with the highest significance level measured by t-/z-statistics) negatively. As for MHHInc and MHHIncPer, however, the correlations between MHHInc and deaths were observed statistically significant in OLS and SEM, but the correlating directions were inconsistent between the two models (positive in OLS and negative in SEM); MHHIncPer was found to be significantly associated with deaths in only SEM and the direction was positive.

[1] The explanatory variables have different units with different value ranges, hence their coefficients are not comparable; the associated t-statistics (OLS) and z-statistics (SLM and SEM) instead could be compared in terms of the significance level of the associations.

### 3.2.2 Model 2: Static local regression analysis

The (M)GWR derived local spatial heterogeneity of the determinant factors for the COVID-19 cases and deaths is shown in **Table 2** and **Fig. 2**. These figures collectively demonstrate the spatial variability of the local model at the county scale in the contiguous US. Local $R^2$ estimates for both local regression models, i.e., MGWR and GWR, have shown high degrees of spatial agreement. The counties, for which the highest $R^2$ (i.e., $R^2$>0.90) values were computed, have formed a spatially clustered patterns across the country. The high values of local $R^2$ were concentrated over the Wisconsin-Indiana-Michigan region, as well as several parts of Texas, California, Mississippi and Arkansas. The lowest $R^2$ score was found in the Northern and North-Western states (i.e., Montana, Washington, Oregon, Wyoming), Southern states (i.e., New Mexico) and North-East coast region (i.e., North Carolina and Georgia). For COVID-19 deaths, the spatial pattern of high, moderate and low $R^2$ values was found similar to those of the COVID-19 cases. Among the two local spatial regression models, MGWR performed more accurately with higher adjusted $R^2$ values (for cases, $R^2$ = 0.969; for deaths, $R^2$ = 0.962), compared to GWR (for cases, $R^2$ = 0.964; for deaths, $R^2$ = 0.954). Also, AICc values of the MGWR model (for cases, AICc = -434.883; for deaths, AICc = -358.146) were found much lower than GWR (for cases, AICc = 238.888; for deaths, AICc = 230.621) (**Table. 2** and **Fig. 2**).

### 3.2.3 Model 3: Group-wise static local regression analysis

The spatial association between different groups (i.e., crime, demographic, education, employment, ethnicity, health and migration) and COVID-19 cases and deaths were examined and presented in **Fig. 3, Fig. 4**. Among the seven groups, six groups viz. Demography, Crime, Education, Ethnicity, Employment, and Population Migration have shown strong similarities in their spatial pattern of local $R^2$. The high local $R^2$ ($R^2$ = > 0.90) was found in the Southern and South-Western states, mainly Texas, Arizona, California, Utah; in the Eastern USA, i.e. Wisconsin-Michigan-Indiana-Illinois region; and in the tri-state area of Mississippi-Arkansas-Alabama. In contrary, the health factor has exhibited a different association with the COVID-19 numbers. High local associations between the health factor and the COVID-19 cases were found in the Colorado-Utah states and New Hampshire areas. For all groups, low spatial associations were found in Montana, North Dakota, Idaho, Oregon states. Based on the $R^2$ and AICc value, the population migration factor was found to be the most critical components with highest local estimates ($R^2$ = 0.96, AICc = -462.76), followed by education, and crime. A similar spatial association was detected between the explanatory factors and COVID-19 death across the counties. The high local association was found over the South, South-Western USA (Texas, New Mexico, Arizona, California states) and in the East Central states (Wisconsin, Michigan, Indiana, Illinois states). Population and migration factor have explained the maximum model variability ($R^2$ = 0.98 and AICc = -2371.52), followed by Employment, Crime, Health, Ethnicity, Demography and Education (**Table 2**).

### 3.2.4 Model 4: Dynamic local regression analysis

Spatial and temporal association between the final selected variables (i.e., six variables) and COVID-19 cases are presented in **Fig. 5, Fig. 6,** and **Table 3**. Totally ten, i.e. five for cases and five for deaths, local regression models were developed for exploring the local associations between the explanatory factors and COVID-19 counts in each of the five months (March, April, May, June, and July). In March, a high spatial association between explanatory variables and response variable was found high in Texas, New Mexico, Mississippi, Tennessee, Kentucky, Indiana, Illinois, Wisconsin and Michigan region ($R^2$ >= 0.90). In April, in addition to the previous regions, a high spatial association was found in Florida and California. In June and July, Arizona, Nevada, Oregon, Idaho states have exhibited a high spatial association characterised by high local $R^2$ values. On the contrary, a low spatial association was observed in Washington, Oregon, Idaho, Montana, North Dakota, South Dakota. For COVID-19 death, the local association has followed a similar pattern as observed for the case. In March, a high spatial association was seen in the Wisconsin and Illinois states. In the later months, the high spatial association was shifted to multiple locations, such as Texas, California, Utah, Idaho, Wyoming region, Arkansas, Mississippi, Tennessee region. On the contrary, a low spatial association was found in the northern (i.e., Montana, North Dakota) and eastern states (i.e., Florida, Georgia, South Carolina). All the dynamic models have demonstrated the superiority of MGWR, as it found to be the best-suited model for local regression analysis throughout the study (**Fig. 5, Fig. 6, Table 3**).

## *3.3 Variable importance*

Relative importance (RI) of the selected variables (final filtered variables, six for cases and six for deaths) measured using the random forest machine learning model were presented in **Fig. 7**. For COVID-19 cases, among the variables, the highest RI was calculated for HBACM (44.31%), followed by DomMig (15.56%), ARSON (12.38%), RIntMig (10.53%),MHHIncPer (5.22%), and MHHInc (3.7%), respectively (**Fig. 7a**). For COVID-19 deaths, the HBAF have explained the maximum variances, and therefore, the highest RI score was computed for HBAF (26.56%), followed by DomMig (13.23%), RDomMig (8.07%), MHHInc (6.84%), RIntMig (5.88%), and MHHIncPer (0.76%), respectively (**Fig. 7b**).

# 4. Discussion

Since the outbreak of COVID-19 starting from Wuhan (China) locally to the global crisis, it has been nearly one year passed, yet the situation remains globally elusive. Among all the countries, the United States (US) is facing the most critical challenge in flattening the curve with more effective and appropriate control measurements. To inform the policy-makers at both national and state levels, understanding the explanatory forces and related confounding factors with spatial patterns is of paramount importance. Studies have done much work of doing so (Mollalo et al., 2020, Sannigrahi et al., 2020a,b). Yet, this could not uncover the full picture since most of the factors change over time, namely being time-variant variables. The present study contributes to forwarding the knowledge of the outbreak by examining a set of factors over space and across time. Specifically, the most relevant variables were teased out from a large group of potential factors for explaining the COVID-19 cases and deaths at the county level, as well as for each month of the five-month study period.

Choosing the best models have always been the crucial point in spatial epidemiological research. Previously, several methodological approaches have been evolved to capture the influence of explanatory variables on response variables in epidemiological study (Bashir et al., 2020; Mollalo et al., 2020). These are Spearman's, Pearson's and Kendall's Correlation Coefficient (Méndez-Arriaga, 2020), ordinary least square regression (Méndez-Arriaga, 2020),

Poisson regression, distributed lag nonlinear model (DLNM) (Andersen et al., 2021; Runkle et al., 2020), cluster based analysis (Andersen et al., 2021), spatial lag model (Sun et al., 2020), spatial error model (Sun et al., 2020), are few of them. These models are mainly global models in nature and therefore have proven ineffective to capture the local/spatial pattern between explanatory and response variables.

Notably, the overall regression models revealed that population migration, as indicated by domestic migration and rate of international migration, is highly correlated with the COVID-19 case and death numbers. The move of people across continents internationally appears to be accompanied with high risk of virus spread, as the air traveling means by its nature increases the likelihood of person-to-person COVID-19 transmissions (Zhang et al., 2020). Given this evidence, airflight restrictions could be effective in undermining the virus spread, which is in line with the conclusion of positive associations between travel restrictions and COVID-19 spread from previous findings (Christidis & Christodoulou 2020), although this involves trade-offs between air-transporting public health and social-economics risks (Cotfas et al., 2020). The other population moving variable, domestic migration, is found to be negatively related to numbers of both cases and deaths, which might be because the redistribution of population from high density areas (e.g., megacities) to low population density areas (e.g. mountainous suburban) could diffuse infected people while decreasing the frequency of person-to-person contact. A study found that residents from New York City, especially those in high wealth status, tended to flee the city to lower physical exposure to COVID-19 (Coven & Gupta, 2020). It should be noted that this relationship is based on the overall regression model, lacking heterogeneity over space and time. Socioeconomically, median household income at the county level positively relates to COVID-19 spread as it indicates the larger cities and higher population densities with more burden of virus transmissions.

Interestingly, when viewing different time periods (monthly from March to July) as revealed from the dynamic local regression analysis, there exists high spatial heterogeneity in how the explanatory variables associate with COVID-19 cases and deaths, and such heterogeneity is dynamic over time, which is also supported by the better performance of MGWR than that of GWR (**Fig. 5** and **Fig. 6**). At the early stage of the COVID-19 outbreak (mainly in March), the associations between the potential factors and the infected numbers in most regions have not been well manifested except for the Chicago-centred Great Lake region and the Tennessee-Arkansas-Mississippi region (**Fig. 5c**). However, since April, several prominent hotspots of such correlations have been discovered including the states of California and Florida as well as many middle east regions (**Fig. 5d, g, h, k**). These regions identified as hotspots have characteristics of high population densities and hence the outbreak outcomes are more likely to be explained by the selected factors, particularly the migration-related variables of domestic migration behaviours. This implication again demonstrates the importance of controlling people mobility as effective measures to combat the virus spread by the government in high populated states (Badr et al., 2020), as those actions taken in other countries including China (Kraemer et al., 2020). In terms of COVID-19 deaths, the spatial patterns of the modelling outcomes also began to exhibit high explanatory powers over large scales after April and remain stable during April-July, covering most of the contiguous US (except for a few regions such as northern California and northern New York). These results confirm that the selected factors of migration, household economic status etc. could well understand the deaths caused by COVID-19 across counties and states during the study period.

Although the US is equipped with best healthcare facilities in the world, the high-level response to the pandemic has been argued as inadequate and leading to “surprisingly” resurgence of COVID-19 cases in e.g. California[2]. Currently, due to the unavailability of a widespread vaccine, the most effective measures to protect from virus spread and minimise exposure risk are still keeping social distances, wearing masks, and high frequency of washing hands (Badr et al., 2020). At the state level, local governments have been sufficiently vigilant to anticipate the situations and have taken preventive and protective measures (e.g. implementing anti-contagion policies) beyond federal guidance to minimize the potential damage. These government-imposed containment policies include, for instance, large event bans, school closures, and mandating social distances, which could reduce the growth of new cases (Courtemanche et al., 2020). State travel restrictions as well as quarantine rules for out-of-state visitors have been put into practices by many states. Educational institutions transferred from in-person classes to online meetings, or otherwise designed protocols specifying different categories of students/staff/faculty members, regular testing, restricted public room usages, etc.

However, effort has been regarded as seemingly being put in vein based on the possible rebounding trend of newly found cases[3]. Given the critics based on the fact that the US has the number of cases far more than any other country, policy-makers have been placed on a verge of taking critically adaptive and learning actions by referring to successful examples. China, the world's second largest economy following the US, has put tremendous resources for controlling virus spread (primarily through city lockdown), which was reported as effective as potentially prevented hundreds of thousands of cases outside Hubei province (WHO, 2020). Challenges such as those rooted in difference in political systems are admittedly persistent when learning from the way in which China respond to the virus crisis, yet quick actions as the Chinese government has taken should be undoubtedly encouraged as the priority by other countries (Kupferschmidt & Cohen, 2020). With more evidence accumulated for testing the underlying forces of COVID-19 spread, it is urgent to call for taking serious and sophisticated consideration by the federal government of socioeconomics and demographics especially population migration at the county/state level in addition to physical protection at the individual level. Without taking these temporally and spatially dynamic factors into account, the COVID-19 mitigation outcomes and the future of public health of the country in response to the pandemic would remain uncertain and risky.

[2] Website: https://www.latimes.com/opinion/story/2020-07-02/u-s-was-perfectly-equipped-to-beat-coronavirus-federal-government-failed

[3] Websites: 1) https://www.cnn.com/videos/politics/2020/04/12/anthony-fauci-polls-november-rebound-jake-tapper-sotu-vpx.cnn; 2) https://coronavirus.jhu.edu/testing/individual-states

The findings in the present studies are generally in agreement with previous investigations while not only adding values to the existing knowledge of COVID-19 spread in the US but also providing implications for combating the crisis worldwide. Consistent with what have been previously found, several factors including climatic, environmental, socio-economic, demographic, etc. have played key role in determining the overall casualties caused by COVID-19 across the countries. For example, Bashir et al., (2020) found that minimum temperature and average temperature are strongly correlated with the spread of COVID-19 in New York city. Apart from that, specific humidity was found positively related with COVID-19 in four US cities – New Orleans, LA; Albany, GA; Chicago, IL; Seattle, WA (Runkle et al., 2020). Different socio-economic factors, such as median household income equality, percentage of nurse participation and percentage of black people, were also found as determining factors of COVID-19 casualties (Mollalo et al., 2020). In addition, demographic profile of the health care professional i.e., over 55 years old population, was found substantially correlated with this disease (Dowd et al., 2020). Economic profile of the communities including unemployed population and existence of socio-economic disparities, was also found as one of the key regulating factors of COVID-19 casualties in the USA. The present study, however, has not found any significant relationship between climate, air pollution and COVID-19 case or deaths (**Fig. S2**). This finding is in line with the observation of Mollalo et al., 2020.

The present study had explored the local and global associations between the explanatory factors and COVID-19 casualties at the county scale in the contiguous US. Though this study had adopted many relevant approaches and methods to allow more accurate model estimates, which can further be used as a reference for similar research interest and policy design, still, the present research is not free from unavoidable uncertainties and biases that exist both in parameter approximation and model design. Cumulated COVID-19 deaths and cases were used as a dependent variable in the spatial models. Though, we have considered the latest COVID-19 counts (COVID-19 data from January 22 to July 26, 2020, was collected) for the modelling, still, there is high chance to have different estimates if the proposed models are performed considering different time frame datasets. To clearly understand this uncertainty, we compared our modelled estimates with Mollalo et al. observations. This study (Mollalo et al. ) have conducted the analysis considering 90 days of aggregated COVID-19 data. While, in the present research, we have considered 348 variables and sorted out 6 and 5 final uncorrelated variables for COVID-19 cases and deaths, respectively, after considering nearly 184 days of data (both aggregated and daily COVID-19 counts were considered). The final filtered variables identified in our study has not matched perfectly with Mollalo et al. estimation. This could be due to the difference in time frame taken between Mollalo et al. (90 days of COVID-19 data) and our study (184 days of COVID-19 data). Moreover, in our study, we have considered seven factors, i.e. crime, demography, education, ethnicity, employment, health, and population & migration factors, for the modelling and subsequent interpretation. The causal effects of the other factors, such as the lockdown date, the strictness of lockdown (partial or complete), restrictions on social gathering and human mobility, etc. have not been explored in the present research. This could be an issue for future research.

## Conclusion

The present research aimed to explore the local and global associations between the explanatory factors and COVID-19 counts in the contiguous US using various local and global spatial regression models and machine learning algorithm. For capturing the time varying effects of the causal factors on COVID-19 counts, several dynamic local parsimonious models have been conceptualised. Among the confounding factors, crime, income, and migration were found to be strongly associated with COVID-19 casualties, and hence explained the maximum model variances. Both global and local associations among the parameters were varied highly over space and could change across time. This spatial variability of the model estimates exhibit the varied behaviour of the explanatory factors and COVID-19 incidences at the county scale. To inform policy-makers at the nation and state levels, understanding the explanatory forces and related confounding factors with spatial patterns is of paramount importance. Therefore, the present study could be a reference for future spatial epidemiological research and informed decision making in the case of crisis.

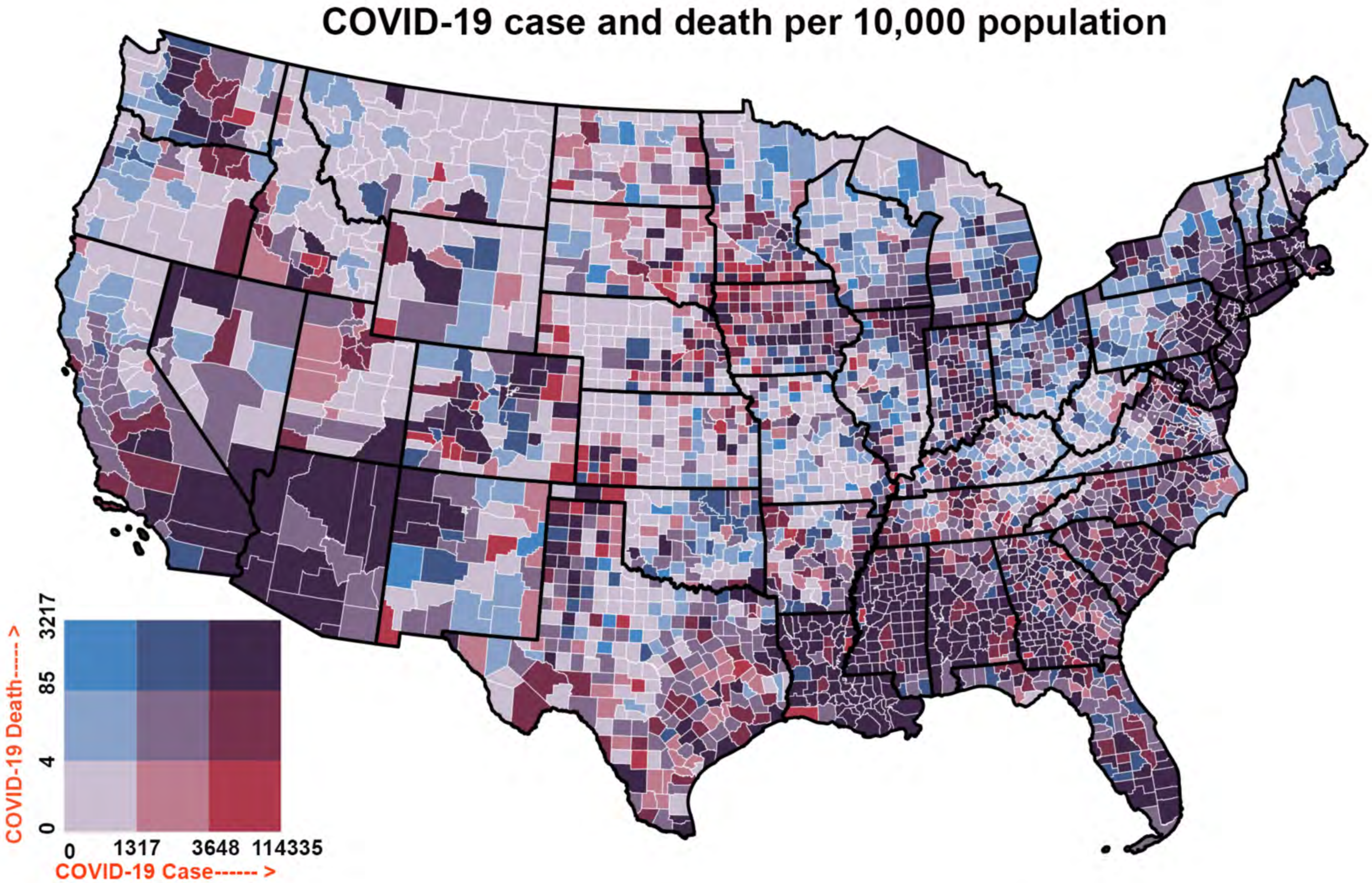

Fig. 1

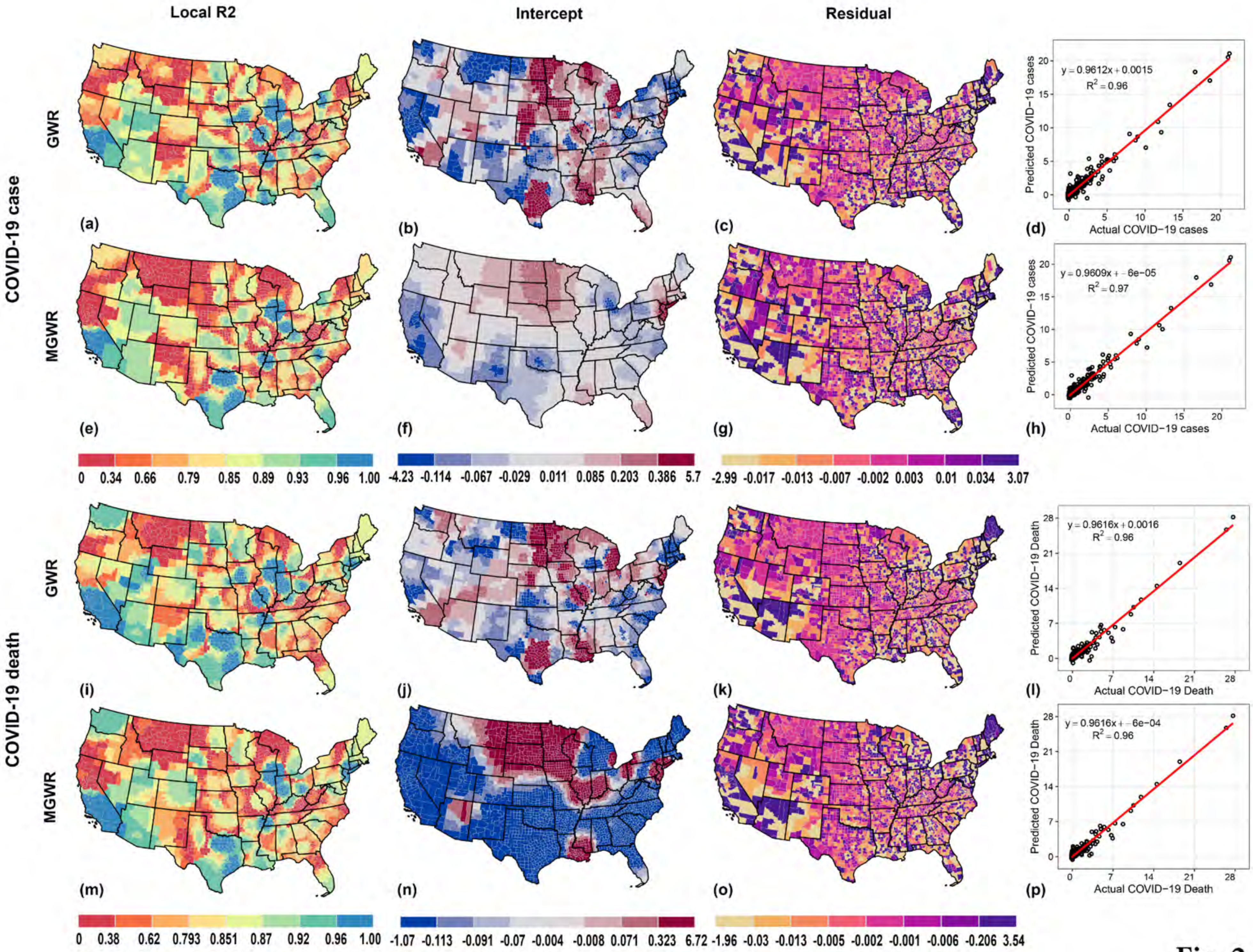

Local R2
Intercept
Residual
COVID-19 case
GWR
MGWR
COVID-19 death
GWR
MGWR
(a)
(b)
(c)
(d)
(e)
(f)
(g)
(h)
(i)
(j)
(k)
(l)
(m)
(n)
(o)
(p)
y = 0.9612x + 0.0015
R2 = 0.96
y = 0.9609x + -6e-05
R2 = 0.97
y = 0.9616x + 0.0016
R2 = 0.96
y = 0.9616x + -6e-04
R2 = 0.96
Predicted COVID-19 cases
Actual COVID-19 cases
Predicted COVID-19 Death
Actual COVID-19 Death
0 0.34 0.66 0.79 0.85 0.89 0.93 0.96 1.00
-4.23 -0.114 -0.067 -0.029 0.011 0.085 0.203 0.386 5.7
-2.99 -0.017 -0.013 -0.007 -0.002 0.003 0.01 0.034 3.07
0 0.38 0.62 0.793 0.851 0.87 0.92 0.96 1.00
-1.07 -0.113 -0.091 -0.07 -0.004 -0.008 0.071 0.323 6.72
-1.96 -0.03 -0.013 -0.005 -0.002 -0.001 -0.006 -0.206 3.54

Fig. 2

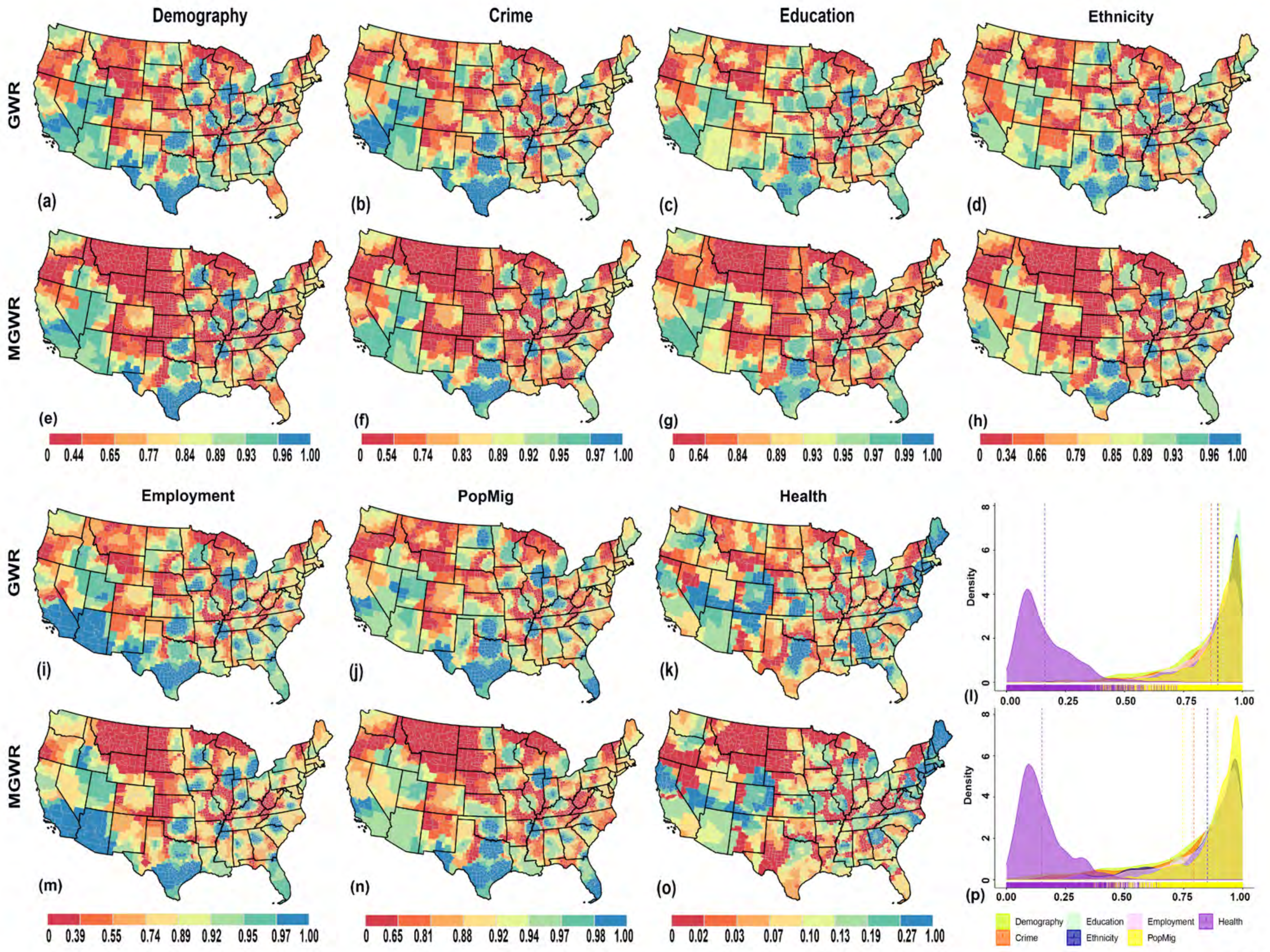

Demography
Crime
Education
Ethnicity
GWR
MGWR
(a)
(b)
(c)
(d)
(e)
(f)
(g)
(h)
0 0.44 0.65 0.77 0.84 0.89 0.93 0.96 1.00
0 0.54 0.74 0.83 0.89 0.92 0.95 0.97 1.00
0 0.64 0.84 0.89 0.93 0.95 0.97 0.99 1.00
0 0.34 0.66 0.79 0.85 0.89 0.93 0.96 1.00
Employment
PopMig
Health
(i)
(j)
(k)
(l)
(m)
(n)
(o)
(p)
Density
0.00 0.25 0.50 0.75 1.00
0 0.39 0.55 0.74 0.89 0.92 0.95 0.97 1.00
0 0.65 0.81 0.88 0.92 0.94 0.97 0.98 1.00
0 0.02 0.03 0.07 0.10 0.13 0.19 0.27 1.00
Demography
Crime
Education
Ethnicity
Employment
PopMig
Health

Fig. 3

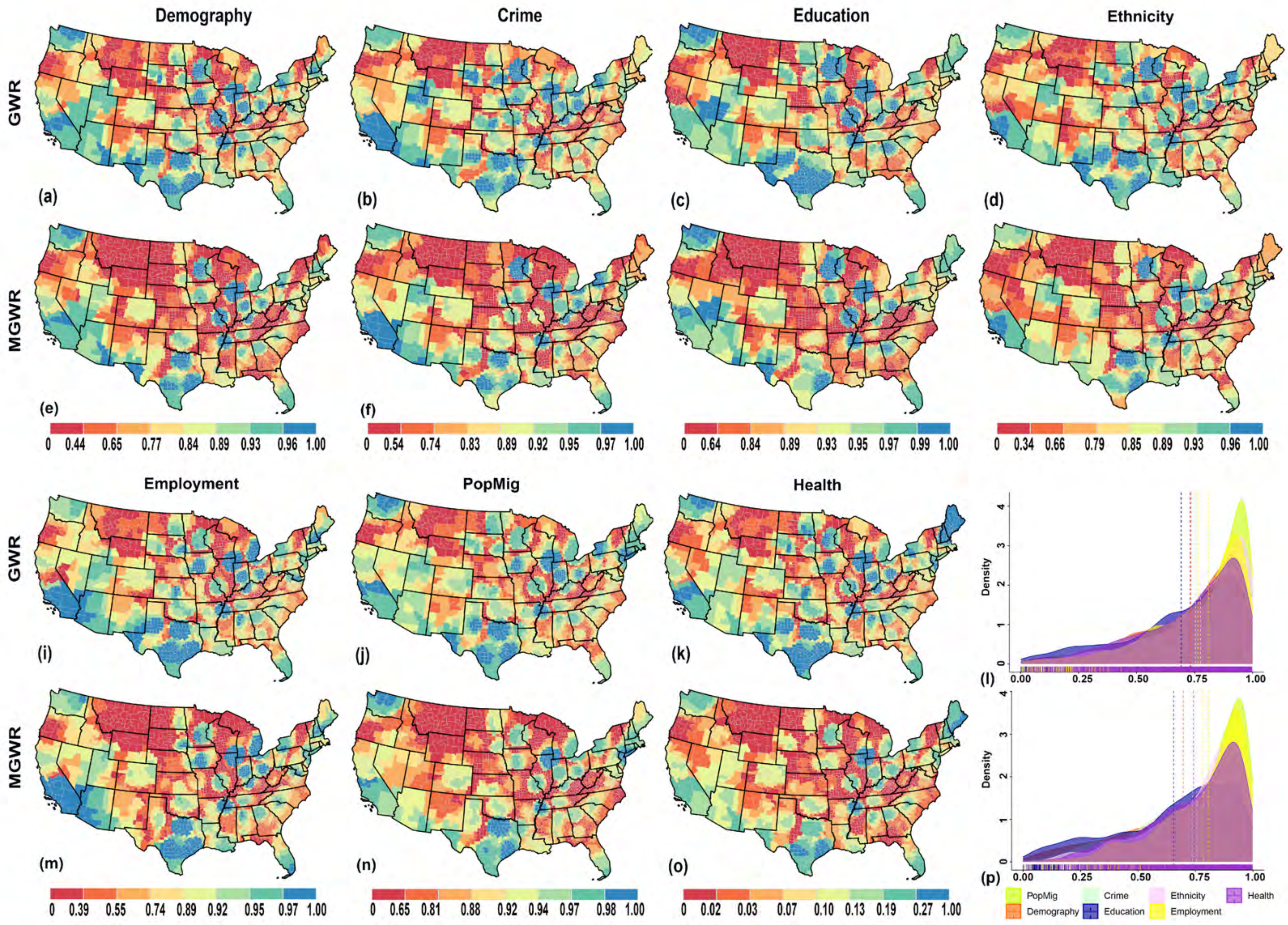
Demography
Crime
Education
Ethnicity
GWR
MGWR
(a)
(b)
(c)
(d)
(e)
(f)
(g)
(h)
0 0.44 0.65 0.77 0.84 0.89 0.93 0.96 1.00
0 0.54 0.74 0.83 0.89 0.92 0.95 0.97 1.00
0 0.64 0.84 0.89 0.93 0.95 0.97 0.99 1.00
0 0.34 0.66 0.79 0.85 0.89 0.93 0.96 1.00
Employment
PopMig
Health
GWR
MGWR
(i)
(j)
(k)
(l)
(m)
(n)
(o)
(p)
Density
0.00 0.25 0.50 0.75 1.00
0 0.39 0.55 0.74 0.89 0.92 0.95 0.97 1.00
0 0.65 0.81 0.88 0.92 0.94 0.97 0.98 1.00
0 0.02 0.03 0.07 0.10 0.13 0.19 0.27 1.00
PopMig
Crime
Ethnicity
Health
Demography
Education
Employment

**Fig. 4**

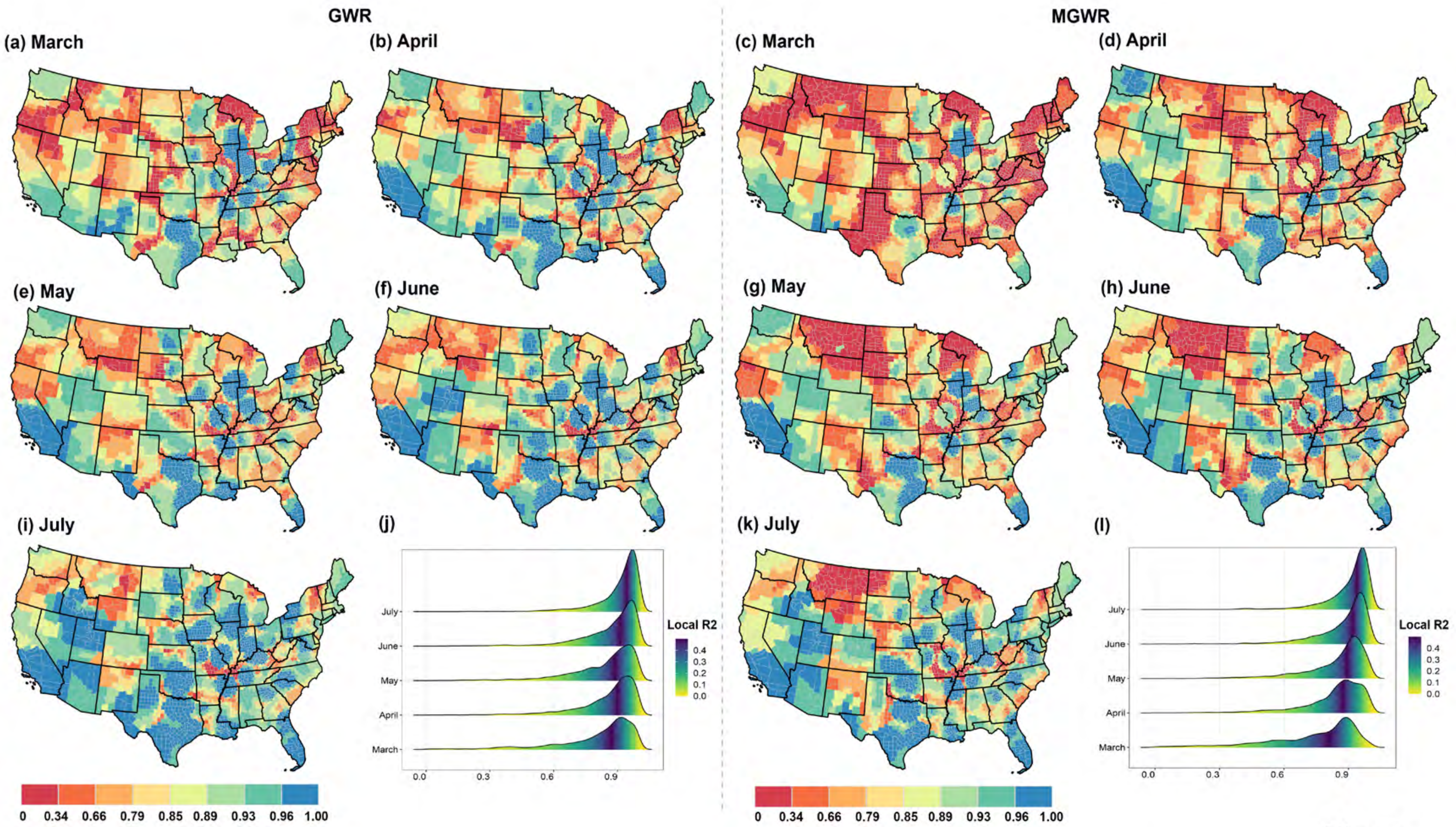

Fig. 5

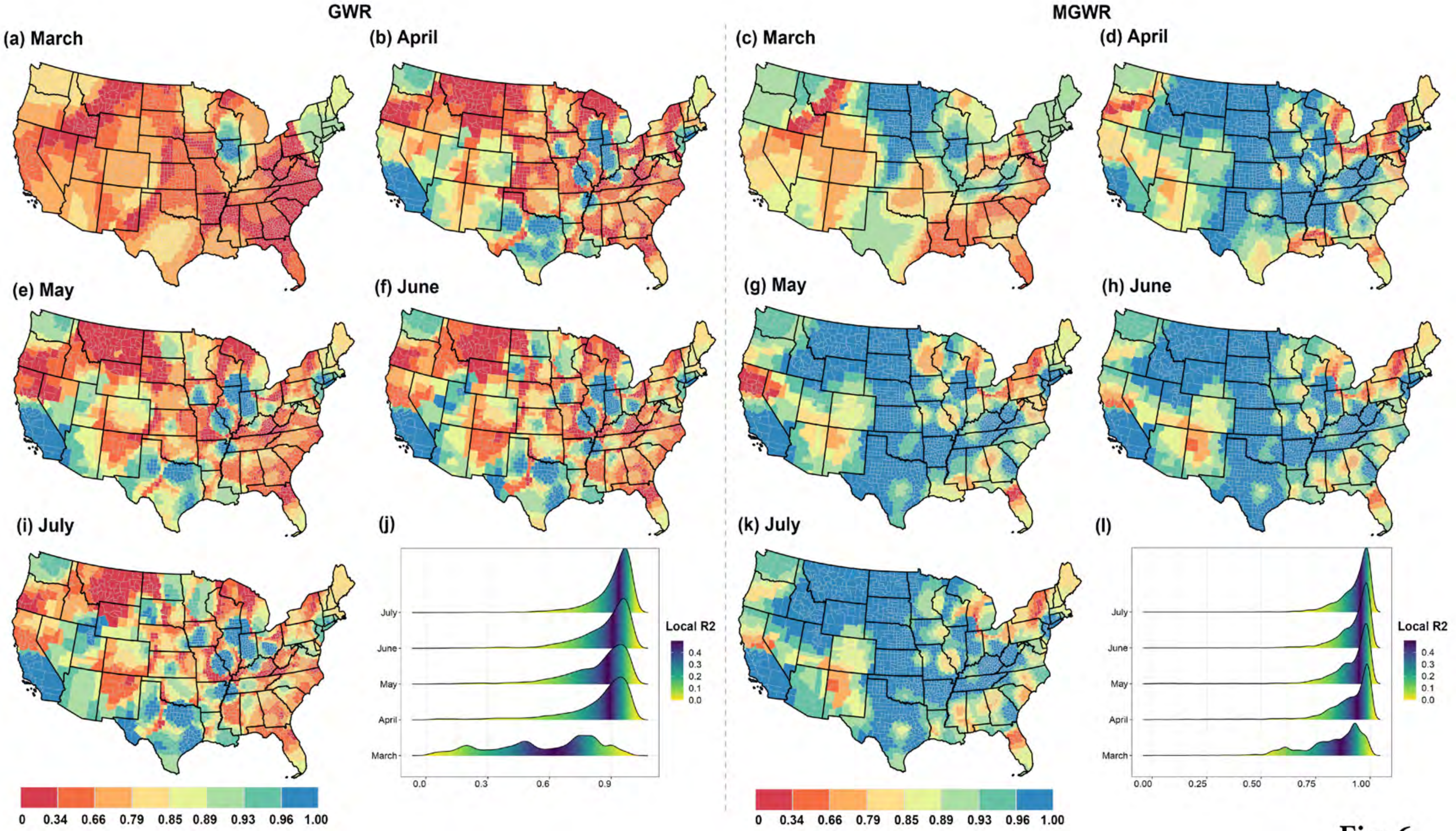

**Fig. 6**

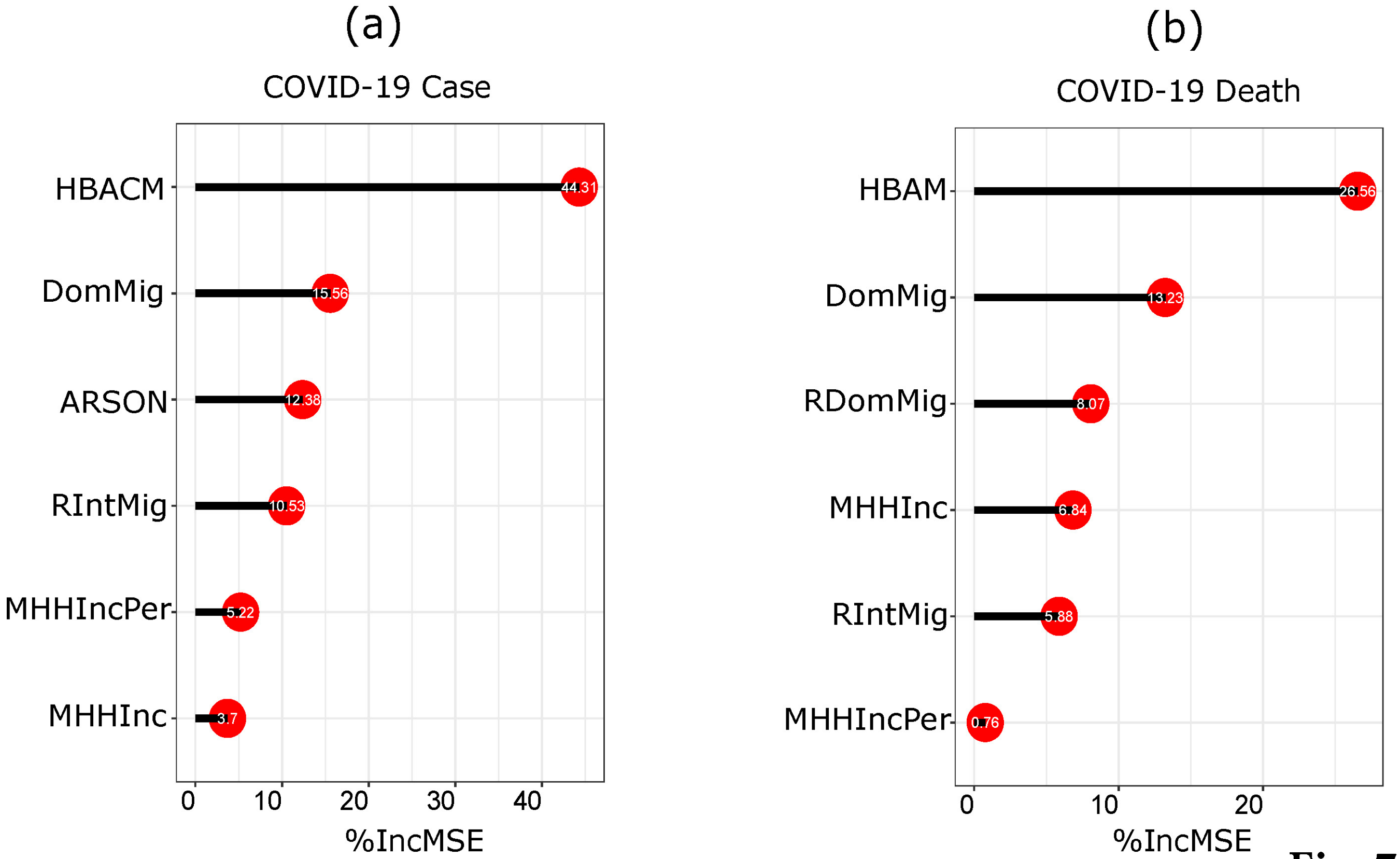

**Fig. 7**

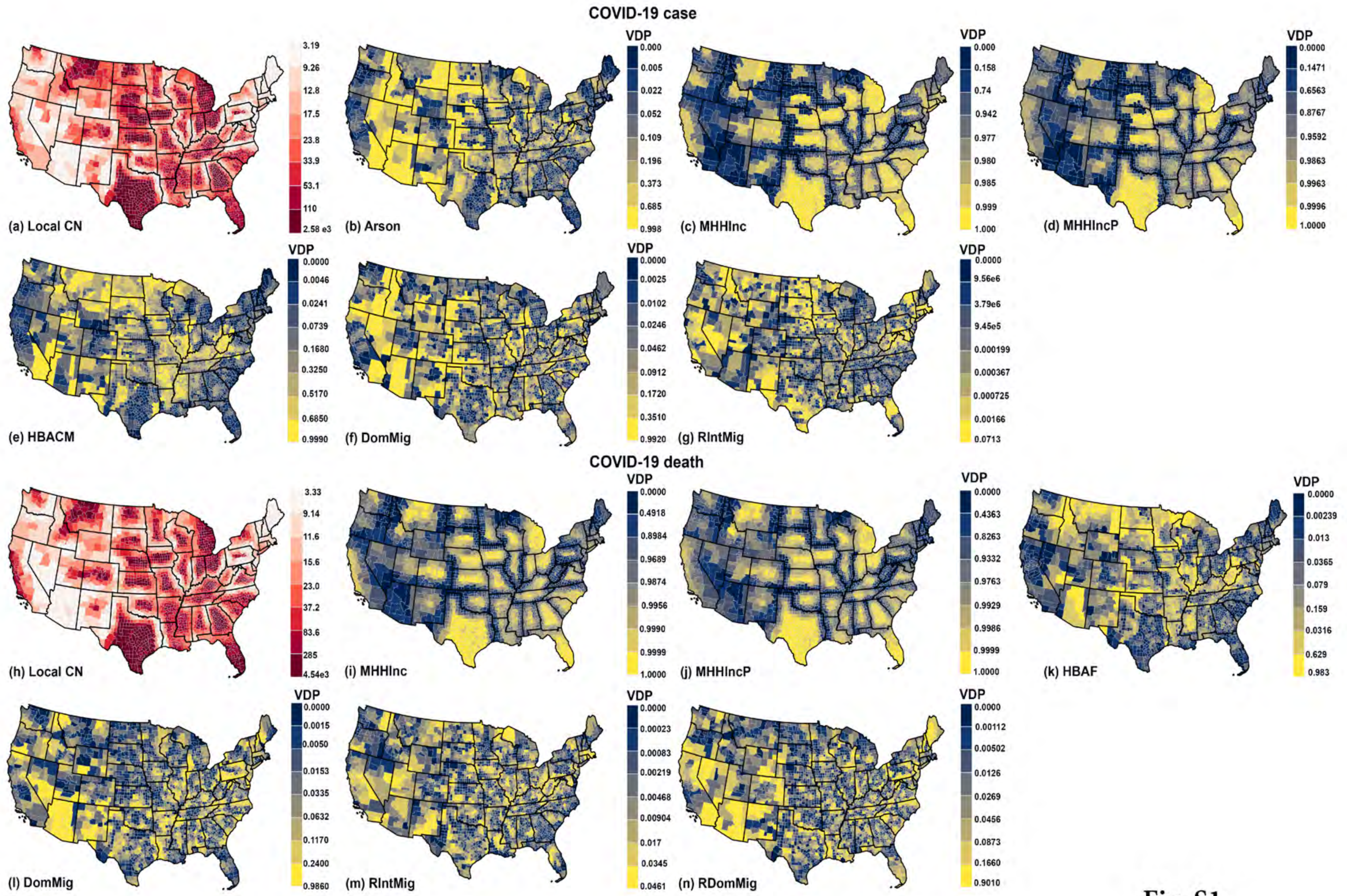

**Fig. S1**

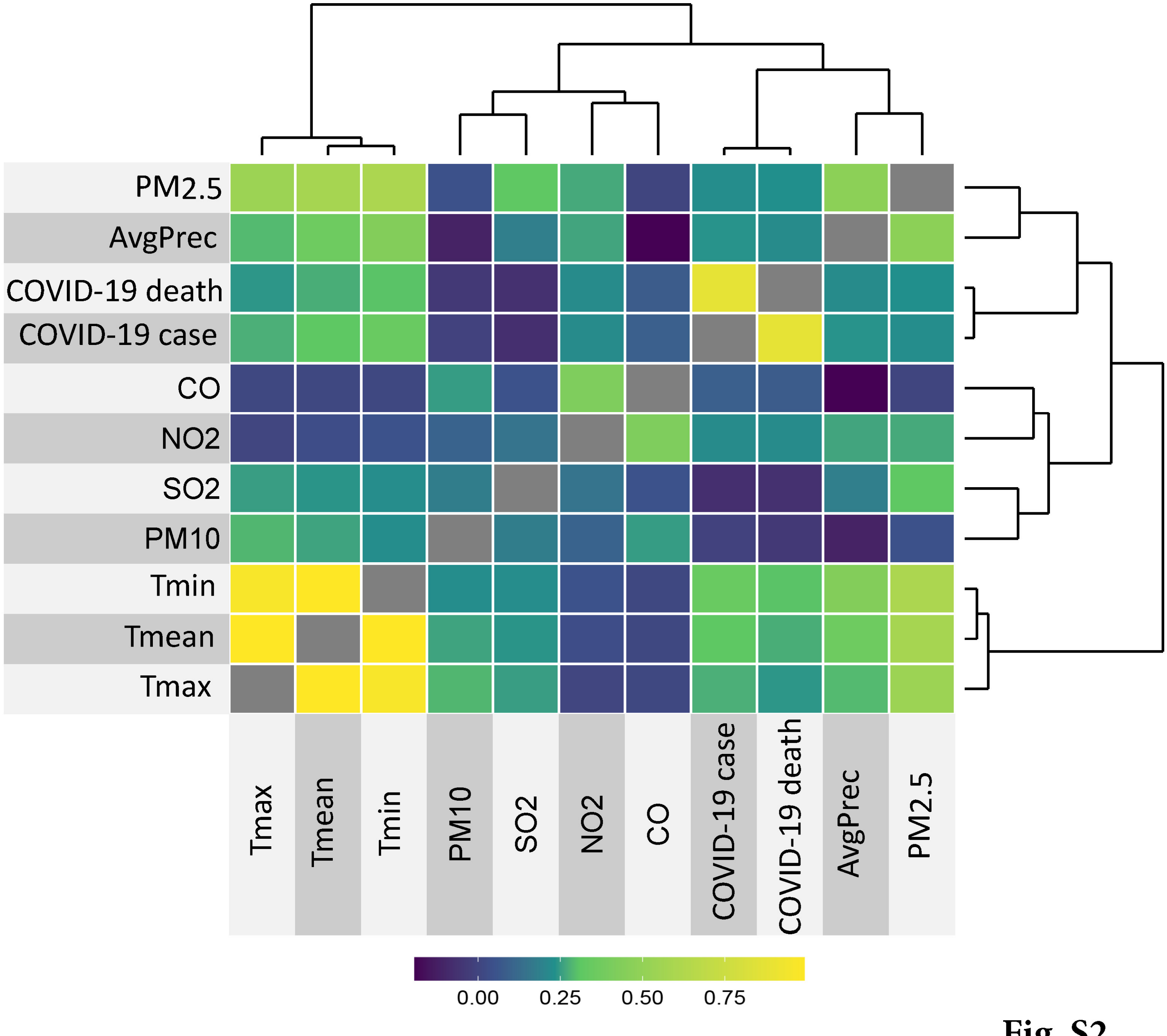

**Fig. S2**

**Table 1** Global regression estimates derived from OLS, SLM, and SEM.

| Variable | **Cases** | | | | | | | | |
|---|---|---|---|---|---|---|---|---|---|
| | **Ordinary Least Square** | | | **Spatial Lag** | | | **Spatial Error** | | |
| | Coefficient | t-Statistic | Probability | Coefficient | z-Statistic | Probability | Coefficient | z-Statistic | Probability |
| Case | --- | --- | --- | 0.34 | 23.34 | 0.00 | --- | --- | --- |
| CONSTANT | -120958 | -8.49 | 0.00 | -54546.7 | -4.1 | 0.00 | -103165 | -6.02 | 0.00 |
| ARSON | 825.14 | 14.88 | 0.00 | 1073.06 | 21.2 | 0.00 | 1291.49 | 23.27 | 0.00 |
| MHHInc | 2.27559 | 5.36 | 0.00 | -0.04 | -0.1 | 0.92 | 2.32 | 3.40 | 0.00 |
| MHHIncPer | 224.50 | 0.78 | 0.43 | 633.62 | 2.43 | 0.02 | 20.76 | 0.05 | 0.96 |
| HBACM | 62.27 | 52.79 | 0.00 | 46.78 | 39.3 | 0.00 | 44.04 | 35.04 | 0.00 |
| DomMig | -22.42 | -20.13 | 0.00 | -21.33 | -20.94 | 0.00 | -22.37 | -20.73 | 0.00 |
| RIntMig | 274.51 | 0.17 | 0.86 | -392.91 | -0.27 | 0.78 | -87.65 | -0.06 | 0.95 |
| Lambda | --- | --- | ---- | --- | --- | --- | 0.55 | 26.04 | 0.00 |
| $R^2$ | | 0.76 | | | 0.80 | | | 0.80 | |
| Adj. $R^2$ | | 0.76 | | | --- | | | --- | |
| F | | 1611.37 | | | --- | | | --- | |
| P | | 0.00 | | | --- | | | --- | |
| AIC | | 83825 | | | 83325.90 | | | 83458.30 | |
| SIC | | 83867.30 | | | 83374.30 | | | 83500.60 | |
| | **Deaths** | | | | | | | | |
| Death | --- | --- | --- | 0.73 | 56.08 | 0.00 | --- | --- | --- |
| CONSTANT | -9136.95 | -5.65 | 0.00 | 332.33 | 0.29 | 0.76 | 1988.94 | 1.02 | 0.30 |
| MHHInc | 0.29 | 6.27 | 0.00 | -0.05 | -1.48 | 0.13 | -0.13 | -1.69 | 0.09 |
| MHHIncPer | -48.05 | -1.50 | 0.13 | 28.54 | 1.28 | 0.19 | 88.23 | 1.83 | 0.06 |
| DomMig | -3.91 | -37.32 | 0.00 | -2.82 | -37.14 | 0.00 | -2.73 | -36.2 | 0.00 |
| RIntMig | 1106.60 | 6.47 | 0.00 | 510.21 | 4.29 | 0.00 | 250.68 | 2.00 | 0.04 |
| RDomMig | 123.34 | 3.78 | 0.00 | 142.51 | 6.28 | 0.00 | 81.82 | 3.02 | 0.00 |
| LAMBDA | --- | --- | --- | ---- | --- | --- | 0.81 | 63.64 | 0.00 |
| $R^2$ | | 0.36 | | | 0.69 | | | 0.69 | |
| Adj. $R^2$ | | 0.36 | | | --- | | | ---- | |
| F | | 349.83 | | | --- | | | --- | |
| P | | 0.00 | | | --- | | | ---- | |
| AIC | | 70200.90 | | | 68324.50 | | | 68429.20 | |
| SIC | | 70237.10 | | | 68366.80 | | | 68465.50 | |

MHHInc – Median household income, MHHIncPer - Median household income percent, DomMig – Domestic Migration, RIntMig – Rate of International Migration, HBACM – Not Hispanic, Black or African American alone or in combination male population,

**Table 2** Group wise GWR and MGWR estimates computed from COVID case and death.

| Factors | R2 | | Adj. R2 | | Adj. alpha (95%) | Adj. critical t value (95%) | AIC | | AICc | | BIC | |
|---|---|---|---|---|---|---|---|---|---|---|---|---|
| | GWR | MGWR | GWR | MGWR | GWR | GWR | GWR | MGWR | GWR | MGWR | GWR | MGWR |
| **Cases** | | | | | | | | | | | | |
| Crime | 0.96 | 0.953 | 0.954 | 0.95 | 0 | 3.577 | -333.214 | -272.611 | -197.511 | -241.079 | 2240.615 | 1014.878 |
| Demography | 0.935 | 0.93 | 0.927 | 0.925 | 0 | 3.612 | 989.264 | 973.424 | 1065.608 | 1002.018 | 2954.972 | 2201.44 |
| Education | 0.961 | 0.958 | 0.955 | 0.953 | 0 | 3.572 | -395.974 | -388.113 | -265.737 | -316.194 | 2129.215 | 1522.801 |
| Employment | 0.96 | 0.963 | 0.953 | 0.955 | 0 | 3.54 | -220.113 | -362.583 | -33.805 | -158.134 | 2758.259 | 2744.779 |
| Ethnicity | 0.953 | 0.952 | 0.946 | 0.946 | 0 | 3.572 | 136.66 | 81.321 | 266.864 | 171.804 | 2661.55 | 2211.106 |
| Health | 0.412 | 0.439 | 0.332 | 0.398 | 0 | 3.542 | 7915.288 | 7450.479 | 8017.685 | 7482.009 | 10172.44 | 8737.93 |
| PopMig | 0.964 | 0.962 | 0.957 | 0.957 | 0 | 3.552 | -469.484 | -574.647 | -263.696 | -462.762 | 2647.13 | 1778.057 |
| All Variables | 0.964 | 0.969 | 0.954 | 0.961 | 0.001 | 3.473 | -133.397 | -737.655 | 238.838 | -434.883 | 3930.356 | 2971.075 |
| **Death** | | | | | | | | | | | | |
| Crime | 0.936 | 0.941 | 0.927 | 0.934 | 0 | 3.566 | 1077.624 | 701.324 | 1202.174 | 782.08 | 3550.934 | 2719.931 |
| Demography | 0.892 | 0.887 | 0.879 | 0.879 | 0 | 3.612 | 2555.903 | 2460.028 | 2632.247 | 2490.847 | 4521.611 | 3733.372 |
| Education | 0.779 | 0.781 | 0.762 | 0.77 | 0 | 3.515 | 4577.562 | 4401.507 | 4612.928 | 4417.641 | 5938.388 | 5331.019 |
| Employment | 0.948 | 0.953 | 0.938 | 0.944 | 0 | 3.54 | 646.627 | 334.34 | 832.936 | 538.789 | 3624.999 | 3441.702 |
| Ethnicity | 0.925 | 0.926 | 0.914 | 0.918 | 0 | 3.546 | 1542.145 | 1303.461 | 1647.646 | 1361.208 | 3831.098 | 3025.062 |
| Health | 0.936 | 0.939 | 0.929 | 0.932 | 0 | 3.607 | 909.58 | 755.337 | 982.54 | 828.208 | 2833.558 | 2678.189 |
| PopMig | 0.98 | 0.98 | 0.975 | 0.977 | 0 | 3.549 | -2090.17 | -2484.22 | -1759.22 | -2371.52 | 1768.129 | -123.531 |
| All Variables | 0.964 | 0.97 | 0.954 | 0.962 | 0.001 | 3.472 | -138.548 | -731.431 | 230.621 | -358.146 | 3910.451 | 3337.357 |

**Table 3** Month wise GWR and MGWR estimates for case and death.

| Months | Cases | | | | | | | | | | | |
|---|---|---|---|---|---|---|---|---|---|---|---|---|
| | $R^2$ | | Adj. $R^2$ | | Adj. alpha (95%) | Adj. critical t value (95%) | AIC | | AICc | | BIC | |
| | GWR | MGWR | GWR | MGWR | GWR | GWR | GWR | MGWR | GWR | MGWR | GWR | MGWR |
| March | 0.886 | 0.887 | 0.858 | 0.87 | 0.001 | 3.447 | 3290.311 | 2854.937 | 3588.543 | 2974.945 | 6971.588 | 5284.201 |
| April | 0.931 | 0.944 | 0.914 | 0.932 | 0.001 | 3.447 | 1719.785 | 943.653 | 2018.018 | 1169.966 | 5401.063 | 4195.497 |
| May | 0.953 | 0.962 | 0.941 | 0.953 | 0.001 | 3.447 | 541.379 | -141.82 | 839.612 | 158.405 | 4222.657 | 3550.403 |
| Jun | 0.966 | 0.971 | 0.956 | 0.964 | 0.001 | 3.473 | -332.63 | -939.287 | 40.276 | -631.086 | 3731.492 | 2796.307 |
| July | 0.974 | 0.976 | 0.966 | 0.97 | 0 | 3.49 | -1077.61 | -1533.93 | -647.896 | -1217.58 | 3246.505 | 2245.248 |
| | **Death** | | | | | | | | | | | |
| March | 0.855 | 0.912 | 0.844 | 0.897 | 0.002 | 3.161 | 3262.754 | 2180.349 | 3297.051 | 2343.071 | 4602.697 | 4977.473 |
| April | 0.957 | 0.965 | 0.945 | 0.957 | 0.001 | 3.472 | 371.277 | -470.549 | 741.026 | -224.624 | 4420.234 | 2905.75 |
| May | 0.959 | 0.969 | 0.95 | 0.961 | 0.001 | 3.42 | -11.612 | -677.245 | 229.337 | -360.743 | 3333.698 | 3102.754 |
| Jun | 0.963 | 0.969 | 0.953 | 0.96 | 0.001 | 3.472 | -120.738 | -586.839 | 249.011 | -212.922 | 3928.219 | 3482.117 |
| July | 0.962 | 0.966 | 0.951 | 0.957 | 0.001 | 3.472 | 37.51 | -372.678 | 407.259 | -6.848 | 4086.467 | 3657.341 |

**Table. 4** Changes in Local $R^2$ values in different months.

| $R^2$ range | Case | | | | | | | | | |
|---|---|---|---|---|---|---|---|---|---|---|
| | March | | April | | May | | June | | July | |
| | GWR | MGWR | GWR | MGWR | GWR | MGWR | GWR | MGWR | GWR | MGWR |
| 0 - 0.34 | 379 | 851 | 167 | 362 | 104 | 311 | 76 | 214 | 41 | 145 |
| 0.34 - 0.66 | 384 | 553 | 300 | 413 | 366 | 403 | 219 | 321 | 133 | 185 |
| 0.66 - 0.79 | 391 | 449 | 378 | 462 | 541 | 410 | 420 | 384 | 224 | 304 |
| 0.79 - 0.85 | 347 | 277 | 356 | 409 | 306 | 320 | 352 | 313 | 243 | 281 |
| 0.85 - 0.89 | 349 | 341 | 337 | 397 | 333 | 321 | 303 | 308 | 263 | 299 |
| 0.89 - 0.93 | 465 | 358 | 492 | 419 | 452 | 511 | 428 | 401 | 447 | 394 |
| 0.93 - 0.96 | 347 | 161 | 427 | 253 | 387 | 428 | 438 | 468 | 519 | 509 |
| 0.96 - 1.00 | 447 | 119 | 652 | 394 | 620 | 405 | 873 | 700 | 1239 | 992 |
| | Death | | | | | | | | | |
| 0 - 0.34 | 878 | 63 | 538 | 80 | 414 | 45 | 358 | 17 | 264 | 11 |
| 0.34 - 0.66 | 892 | 414 | 642 | 102 | 680 | 62 | 575 | 60 | 512 | 45 |
| 0.66 - 0.79 | 643 | 501 | 575 | 178 | 559 | 177 | 547 | 130 | 511 | 122 |
| 0.79 - 0.85 | 367 | 420 | 351 | 325 | 302 | 270 | 348 | 201 | 418 | 227 |
| 0.85 - 0.89 | 57 | 316 | 243 | 319 | 304 | 321 | 253 | 287 | 262 | 260 |
| 0.89 - 0.93 | 186 | 449 | 192 | 421 | 281 | 391 | 330 | 376 | 325 | 411 |
| 0.93 - 0.96 | 31 | 504 | 193 | 336 | 233 | 379 | 280 | 454 | 328 | 475 |
| 0.96 - 1.00 | 55 | 442 | 375 | 1348 | 336 | 1464 | 418 | 1584 | 489 | 1558 |

# Table S1 List of all variables (348) used in the study.

| Group | Data variable | Description |
|---|---|---|
| Pollution | PM25_2019 | Particulate Matter 2.5/ug m3 in 2019 |
| | PM10_2019 | Particulate Matter 10/ug m3 in 2019 |
| | SO2_2019 | Sulphur Dioxide/ug m3 in 2019 |
| | NO2_2019 | Nitrogen Dioxide/ug m3 in 2019 |
| | CO_2019 | Carbon Monoxide/mg m3 in 2019 |
| Climate | Precipitation / inch | Precipitation/inch for 2019 |
| | Temp AVG / F | Average temperature in 2019 |
| | Temp Min / F | Minimum temperature in 2019 |
| | Temp Max / F | Maximum temperature in 2019 |
| Demographics | Total_Male | Total number of males |
| | Total_Female | Total number of females |
| | Total_age0to17 | Total number of people between ages 0 to 17 |
| | Male_age0to17 | Total number of males between ages 0 to 17 |
| | Female_age0to17 | Total number of females between ages 0 to 17 |
| | Total_age18to64 | Total number of people between ages 18 to 64 |
| | Male_age18to64 | Total number of males between ages 18 to 64 |
| | Female_age18to64 | Total number of females between ages 18 to 64 |
| | Total_age65plus | Total number of people with age >65 |
| | Male_age65plus | Total number of males with age >65 |
| | Female_age65plus | Total number of females with age >65 |
| | Total_age85plusr | Total number of people with age >85 |
| | Male_age85plusr | Total number of males with age >85 |
| | Female_age85plusr | Total number of females with age >85 |
| | pop_density | Population density as of 2018 |
| Education | Less than a high school diploma 2014-18 | Number of adults who do not have a high school diploma |
| | High school diploma only 2014-18 | Number of adults who just have a high school diploma |
| | Some college or associate's degree 2014-18 | Number of adults with a college degree below bachelor's |
| | Bachelor's degree or higher 2014-18 | Number of adults who have at least a bachelor's degree |
| | Percent of adults with less than a high school diploma 2014-18 | Percentage of adults who do not have a high school diploma |
| | Percent of adults with a high school diploma only 2014-18 | Percentage of adults who just have a high school diploma |
| | Percent of adults completing some college or associate's degree 2014-18 | Percentage of adults with a college degree below bachelor's |
| | Percent of adults with a bachelor's degree or higher 2014-18 | Percentage of adults who have at least a bachelor's degree |
| | POVALL_2018 | Estimate of people of all ages in poverty 2018 |
| | CI90LBAll_2018 | 90% confidence interval lower bound of estimate of people of all ages in poverty 2018 |
| | CI90UBALL_2018 | 90% confidence interval upper bound of estimate of people of all ages in poverty 2018 |
| | PCTPOVALL_2018 | Estimated percent of people of all ages in poverty 2018 |
| | CI90LBALLP_2018 | 90% confidence interval lower bound of estimate of percent of people of all ages in poverty 2018 |
| | CI90UBALLP_2018 | 90% confidence interval upper bound of estimate of percent of people of all ages in poverty 2018 |
| | POV017_2018 | Estimate of people age 0-17 in poverty 2018 |
| | CI90LB017_2018 | 90% confidence interval lower bound of estimate of people age 0-17 in poverty 2018 |
| | CI90UB017_2018 | 90% confidence interval upper bound of estimate of people age 0-17 in poverty 2018 |
| Education | PCTPOV017_2018 | Estimated percent of people age 0-17 in poverty 2018 |
| | CI90LB017P_2018 | 90% confidence interval lower bound of estimate of percent of people age 0-17 in poverty 2018 |
| | CI90UB017P_2018 | 90% confidence interval upper bound of estimate of percent of people age 0-17 in poverty 2018 |
| | POV517_2018 | Estimate of related children age 5-17 in families in poverty 2018 |
| | CI90LB517_2018 | 90% confidence interval lower bound of estimate of related children age 5-17 in families in poverty 2018 |
| | CI90UB517_2018 | 90% confidence interval upper bound of estimate of related children age 5-17 in families in poverty 2018 |
| | PCTPOV517_2018 | Estimated percent of related children age 5-17 in families in poverty 2018 |
| | CI90LB517P_2018 | 90% confidence interval lower bound of estimate of percent of related children age 5-17 in families in poverty 2018 |
| | CI90UB517P_2018 | 90% confidence interval upper bound of estimate of percent of related children age 5-17 in families in poverty 2018 |
| | MEDHHINC_2018 | Estimate of median household income 2018 |
| | CI90LBINC_2018 | 90% confidence interval lower bound of estimate of median household income 2018 |
| | CI90UBINC_2018 | 90% confidence interval upper bound of estimate of median household income 2018 |
| Employment and median household income | Civilian_labor_force_2018 | Civilian labor force annual average |
| | Employed_2018 | Number employed annual average |
| | Unemployed_2018 | Number unemployed annual average |
| | Unemployment_rate_2018 | Unemployment rate |
| | Median_Household_Income_2018 | Estimate of Median household Income, 2018 |
| | Med_HH_Income_Percent_of_State_Total_2018 | County Household Median Income as a percent of the State Total Median Household Income, 2018 |
| Ethnicity | TOT_POP | Total population |
| | TOT_MALE | Total male population |
| | TOT_FEMALE | Total female population |
| | WA_MALE | White alone male population |
| | WA_FEMALE | White alone female population |
| | BA_MALE | Black or African American alone male population |
| | BA_FEMALE | Black or African American alone female population |
| | IA_MALE | American Indian and Alaska Native alone male population |
| | IA_FEMALE | American Indian and Alaska Native alone female population |
| | AA_MALE | Asian alone male population |
| | AA_FEMALE | Asian alone female population |
| | NA_MALE | Native Hawaiian and Other Pacific Islander alone male population |
| | NA_FEMALE | Native Hawaiian and Other Pacific Islander alone female population |
| | TOM_MALE | Two or More Races male population |
| | TOM_FEMALE | Two or More Races female population |
| | WAC_MALE | White alone or in combination male population |
| | WAC_FEMALE | White alone or in combination female population |
| | BAC_MALE | Black or African American alone or in combination male population |
| | BAC_FEMALE | Black or African American alone or in combination female population |
| | IAC_MALE | American Indian and Alaska Native alone or in combination male population |
| | IAC_FEMALE | American Indian and Alaska Native alone or in combination male population |
| Ethnicity | AAC_MALE | Asian alone or in combination male population |
| | AAC_FEMALE | Asian alone or in combination female population |
| | NAC_MALE | Native Hawaiian and Other Pacific Islander alone or in combination male population |
| | NAC_FEMALE | Native Hawaiian and Other Pacific Islander alone or in combination female population |
| | NH_MALE | Not Hispanic male population |
| | NH_FEMALE | Not Hispanic female population |
| | NHWA_MALE | Not Hispanic, White alone male population |
| | NHWA_FEMALE | Not Hispanic, White alone female population |
| | NHBA_MALE | Not Hispanic, Black or African American alone male population |
| | NHBA_FEMALE | Not Hispanic, Black or African American alone female population |
| | NHIA_MALE | Not Hispanic, American Indian and Alaska Native alone male population |
| | NHIA_FEMALE | Not Hispanic, American Indian and Alaska Native alone female population |
| | NHAA_MALE | Not Hispanic, Asian alone male population |
| | NHAA_FEMALE | Not Hispanic, Asian alone female population |
| | NHNA_MALE | Not Hispanic, Native Hawaiian and Other Pacific Islander alone male population |
| | NHNA_FEMALE | Not Hispanic, Native Hawaiian and Other Pacific Islander alone female population |
| | NHTOM_MALE | Not Hispanic, Two or More Races male population |
| | NHTOM_FEMALE | Not Hispanic, Two or More Races female population |
| | NHWAC_MALE | Not Hispanic, White alone or in combination male population |
| | NHWAC_FEMALE | Not Hispanic, White alone or in combination female population |
| | NHBAC_MALE | Not Hispanic, Black or African American alone or in combination male population |
| | NHBAC_FEMALE | Not Hispanic, Black or African American alone or in combination female population |
| | NHIAC_MALE | Not Hispanic, American Indian and Alaska Native alone or in combination male population |
| | NHIAC_FEMALE | Not Hispanic, American Indian and Alaska Native alone or in combination female population |
| | NHAAC_MALE | Not Hispanic, Asian alone or in combination male population |
| | NHAAC_FEMALE | Not Hispanic, Asian alone or in combination female population |
| | NHNAC_MALE | Not Hispanic, Native Hawaiian and Other Pacific Islander alone or in combination male population |
| | NHNAC_FEMALE | Not Hispanic, Native Hawaiian and Other Pacific Islander alone or in combination female population |
| | H_MALE | Hispanic male population |
| | H_FEMALE | Hispanic female population |
| | HWA_MALE | Hispanic, White alone male population |
| | HWA_FEMALE | Hispanic, White alone female population |
| | HBA_MALE | Hispanic, Black or African American alone male population |
| | HBA_FEMALE | Hispanic, Black or African American alone female population |
| | HIA_MALE | Hispanic, American Indian and Alaska Native alone male population |
| | HIA_FEMALE | Hispanic, American Indian and Alaska Native alone female population |
| | HAA_MALE | Hispanic, Asian alone male population |
| | HAA_FEMALE | Hispanic, Asian alone female population |
| | HNA_MALE | Hispanic, Native Hawaiian and Other Pacific Islander alone male population |
| | HNA_FEMALE | Hispanic, Native Hawaiian and Other Pacific Islander alone female population |
| | HTOM_MALE | Hispanic, Two or More Races male population |
| | HTOM_FEMALE | Hispanic, Two or More Races female population |
| | HWAC_MALE | Hispanic, White alone or in combination male population |
| Ethnicity | HWAC_FEMALE | Hispanic, White alone or in combination female population |
| | HBAC_MALE | Hispanic, Black or African American alone or in combination male population |
| | HBAC_FEMALE | Hispanic, Black or African American alone or in combination female population |
| | HIAC_MALE | Hispanic, American Indian and Alaska Native alone or in combination male population |
| | HIAC_FEMALE | Hispanic, American Indian and Alaska Native alone or in combination female population |
| | HAAC_MALE | Hispanic, Asian alone or in combination male population |
| | HAAC_FEMALE | Hispanic, Asian alone or in combination female population |
| | HNAC_MALE | Hispanic, Native Hawaiian and Other Pacific Islander alone or in combination male population |
| | HNAC_FEMALE | Hispanic, Native Hawaiian and Other Pacific Islander alone or in combination female population |
| Healthcare | All Specialties (AAMC) | All Specialties (Assumed proportion to fraction of state population living in county) |
| | Allergy & Immunology (AAMC) | Allergy & Immunology (Assumed proportion to fraction of state population living in county) |
| | Anatomic/Clinical Pathology (AAMC) | Anatomic/Clinical Pathology (Assumed proportion to fraction of state population living in county) |
| | Anesthesiology (AAMC) | Anesthesiology (Assumed proportion to fraction of state population living in county) |
| | Cardiovascular Disease (AAMC) | Cardiovascular Disease (Assumed proportion to fraction of state population living in county) |
| | Child & Adolescent Psychiatry** (AAMC) | Child & Adolescent Psychiatry** (Assumed proportion to fraction of state population living in county) |
| | Critical Care Medicine (AAMC) | Critical Care Medicine (Assumed proportion to fraction of state population living in county) |
| | Dermatology (AAMC) | Dermatology (Assumed proportion to fraction of state population living in county) |
| | Emergency Medicine (AAMC) | Emergency Medicine (Assumed proportion to fraction of state population living in county) |
| | Endocrinology, Diabetes & Metabolism (AAMC) | Endocrinology, Diabetes & Metabolism (Assumed proportion to fraction of state population living in county) |
| | Family Medicine/General Practice (AAMC) | Family Medicine/General Practice (Assumed proportion to fraction of state population living in county) |
| | Gastroenterology (AAMC) | Gastroenterology (Assumed proportion to fraction of state population living in county) |
| | General Surgery (AAMC) | General Surgery (Assumed proportion to fraction of state population living in county) |
| | Geriatric Medicine*** (AAMC) | Geriatric Medicine*** (Assumed proportion to fraction of state population living in county) |
| | Hematology & Oncology (AAMC) | Hematology & Oncology (Assumed proportion to fraction of state population living in county) |
| | Infectious Disease (AAMC) | Infectious Disease (Assumed proportion to fraction of state population living in county) |
| | Internal Medicine (AAMC) | Internal Medicine (Assumed proportion to fraction of state population living in county) |
| | Internal Medicine/Pediatrics (AAMC) | Internal Medicine/Pediatrics (Assumed proportion to fraction of state population living in county) |
| | Interventional Cardiology (AAMC) | Interventional Cardiology (Assumed proportion to fraction of state population living in county) |
| Healthcare | Neonatal-Perinatal Medicine (AAMC) | Neonatal-Perinatal Medicine (Assumed proportion to fraction of state population living in county) |
| | Nephrology (AAMC) | Nephrology (Assumed proportion to fraction of state population living in county) |
| | Neurological Surgery (AAMC) | Neurological Surgery (Assumed proportion to fraction of state population living in county) |
| | Neurology (AAMC) | Neurology (Assumed proportion to fraction of state population living in county) |
| | Neuroradiology (AAMC) | Neuroradiology (Assumed proportion to fraction of state population living in county) |
| | Obstetrics & Gynecology (AAMC) | Obstetrics & Gynecology (Assumed proportion to fraction of state population living in county) |
| | Ophthalmology (AAMC) | Ophthalmology (Assumed proportion to fraction of state population living in county) |
| | Orthopedic Surgery (AAMC) | Orthopedic Surgery (Assumed proportion to fraction of state population living in county) |
| | Otolaryngology (AAMC) | Otolaryngology (Assumed proportion to fraction of state population living in county) |
| | Pain Medicine & Pain Management (AAMC) | Pain Medicine & Pain Management (Assumed proportion to fraction of state population living in county) |
| | Pediatrics** (AAMC) | Pediatrics** (Assumed proportion to fraction of state population living in county) |
| | Physical Medicine & Rehabilitation (AAMC) | Physical Medicine & Rehabilitation (Assumed proportion to fraction of state population living in county) |
| | Plastic Surgery (AAMC) | Plastic Surgery (Assumed proportion to fraction of state population living in county) |
| | Preventive Medicine (AAMC) | Preventive Medicine (Assumed proportion to fraction of state population living in county) |
| | Psychiatry (AAMC) | Psychiatry (Assumed proportion to fraction of state population living in county) |
| | Pulmonary Disease (AAMC) | Pulmonary Disease (Assumed proportion to fraction of state population living in county) |
| | Radiation Oncology (AAMC) | Radiation Oncology (Assumed proportion to fraction of state population living in county) |
| | Radiology & Diagnostic Radiology (AAMC) | Radiology & Diagnostic Radiology (Assumed proportion to fraction of state population living in county) |
| | Rheumatology (AAMC) | Rheumatology (Assumed proportion to fraction of state population living in county) |
| | Sports Medicine (AAMC) | Sports Medicine (Assumed proportion to fraction of state population living in county) |
| | Thoracic Surgery (AAMC) | Thoracic Surgery (Assumed proportion to fraction of state population living in county) |
| | Urology (AAMC) | Urology (Assumed proportion to fraction of state population living in county) |
| | Vascular & Interventional Radiology (AAMC) | Vascular & Interventional Radiology (Assumed proportion to fraction of state population living in county) |
| | Vascular Surgery (AAMC) | Vascular Surgery (Assumed proportion to fraction of state population living in county) |
| | State/Local Government hospital beds per 1000 people (2019) | State/Local Government hospital beds per 1000 people (2019) (Assumed identical to state) |
| | Non-profit hospital beds per 1000 people (2019) | Non-profit hospital beds per 1000 people (2019) (Assumed identical to state) |
| | For-profit hospital beds per 1000 people (2019) | For-profit hospital beds per 1000 people (2019) (Assumed identical to state) |
| | Total hospital beds per 1000 people (2019) | Total hospital beds per 1000 people (2019) (Assumed identical to state) |
| | Total nurse practitioners (2019) | Total nurses (2019) (Assumed proportion to fraction of state population living in county) |
| | Total physician assistants (2019) | Total physical assistants (2019) (Assumed proportion to fraction of state population living in county) |
| Healthcare | Total Hospitals (2019) | Total Hospitals (2019) (Assumed proportion to fraction of state population living in county) |
| | Internal Medicine Primary Care (2019) | Active Internal Medicine Primary Care Physicians (2019) (Assumed proportion to fraction of state population living in county) |
| | Family Medicine/General Practice Primary Care (2019) | Active Family Medicine/General Practice Primary Care Physicians (2019) (Assumed proportion to fraction of state population living in county) |
| | Pediatrics Primary Care (2019) | Active Pediatrics Primary Care Physicians (2019) (Assumed proportion to fraction of state population living in county) |
| | Obstetrics & Gynecology Primary Care (2019) | Active Obstetrics & Gynecology Primary Care Physicians (2019) (Assumed proportion to fraction of state population living in county) |
| | Geriatrics Primary Care (2019) | Active Geriatrics Primary Care Physicians (2019) (Assumed proportion to fraction of state population living in county) |
| | Total Primary Care specialists (2019) | Sum of Internal Medicine, Family Med/General Practice, Pediatrics, OBGYN, and Geriatrics (2019) (Assumed proportion to fraction of state population living in county) |
| | Psychiatry specialists (2019) | Active Psychiatry specialists (2019) (Assumed proportion to fraction of state population living in county) |
| | Surgery specialists (2019) | Active Surgery specialists (2019) (Assumed proportion to fraction of state population living in county) |
| | Anesthesiology specialists (2019) | Active Anesthesiology specialists (2019) (Assumed proportion to fraction of state population living in county) |
| | Emergency Medicine specialists (2019) | Active Emergency Medicine specialists (2019) (Assumed proportion to fraction of state population living in county) |
| | Radiology specialists (2019) | Active Radiology specialists (2019) (Assumed proportion to fraction of state population living in county) |
| | Cardiology specialists (2019) | Active Cardiology specialists (2019) (Assumed proportion to fraction of state population living in county) |
| | Oncology (Cancer) specialists (2019) | Active Oncology (Cancer) specialists (2019) (Assumed proportion to fraction of state population living in county) |
| | Endocrinology, Diabetes, and Metabolism specialists (2019) | Active Endocrinology, Diabetes, and Metabolism specialists (2019) |
| | All Other Specialties specialists (2019) | All Other Specialties specialists (2019) (Assumed proportion to fraction of state population living in county) |
| | Total specialists (2019) | Sum of Psychiatry, Surgery, Anesthesiology, Emergency Med, Radiology, Cardiology, Oncology, Endocrinology, and Other specialists (2019) (Assumed proportion to fraction of state population living in county) |
| | ICU Beds | Number of ICU beds per county |
| Population Estimates | POP_ESTIMATE_2018 | 7/1/2018 resident total population estimate |
| | INTERNATIONAL_MIG_2018 | Net international migration in period 7/1/2017 to 6/30/2018 |
| | DOMESTIC_MIG_2018 | Net domestic migration in period 7/1/2017 to 6/30/2018 |
| | NET_MIG_2018 | Net migration in period 7/1/2017 to 6/30/2018 |
| | R_INTERNATIONAL_MIG_2018 | Net international migration rate in period 7/1/2017 to 6/30/2018 |
| | R_DOMESTIC_MIG_2018 | Net domestic migration rate in period 7/1/2017 to 6/30/2018 |
| | R_NET_MIG_2018 | Net migration rate in period 7/1/2017 to 6/30/2018 |

**Table S2 Diagnostic statistics for spatial dependence**

| | Test | MI/DF | Value | Probability |
|---|---|---|---|---|
| Cases | Moran's I (error) | 0.18 | 16.97 | 0.00 |
| | Lagrange Multiplier (lag) | 1 | 551.98 | 0.00 |
| | Robust LM (lag) | 1 | 286.63 | 0.00 |
| | Lagrange Multiplier (error) | 1 | 283.13 | 0.00 |
| | Robust LM (error) | 1 | 17.77 | 0.00 |
| | Lagrange Multiplier (SARMA) | 2 | 569.76 | 0.00 |
| Deaths | Moran's I (error) | 0.49 | 46.03 | 0.00 |
| | Lagrange Multiplier (lag) | 1 | 2347.05 | 0.00 |
| | Robust LM (lag) | 1 | 327.75 | 0.00 |
| | Lagrange Multiplier (error) | 1 | 2098.74 | 0.00 |
| | Robust LM (error) | 1 | 79.44 | 0.00 |
| | Lagrange Multiplier (SARMA) | 2 | 2426.47 | 0.00 |